%% file: main.tex
\documentclass[sigconf]{acmart}

\usepackage[utf8]{inputenc} 
\usepackage{subfigure}
\usepackage{enumitem}
\AtBeginDocument{%
  \providecommand\BibTeX{{%
    \normalfont B\kern-0.5em{\scshape i\kern-0.25em b}\kern-0.8em\TeX}}}

\usepackage{xspace} 
\newcommand*{\eg}{e.g.,\xspace}
\newcommand*{\ie}{i.e.,\xspace}

\begin{document}

\title[Teaching Responsible Data Science]{Teaching Responsible Data Science: Charting New Pedagogical Territory}

\author{Julia Stoyanovich and Armanda Lewis}
\affiliation{%
  \institution{New York University, NY, USA}
}
\email{[stoyanovich,al861]@nyu.edu}

\input{abstract} 

\maketitle

\input{intro}
\input{experience}
\input{interpret}
\input{objects} 
\input{methods} 
\input{conc}

\balance 

\bibliographystyle{ACM-Reference-Format}
\bibliography{teach}

\end{document}

%% file: abstract.tex
\begin{abstract}
    
Although numerous ethics courses are available, with many focusing specifically on technology and computer ethics, pedagogical approaches employed in these courses rely exclusively on texts rather than on software development or data analysis. Technical students often consider these courses unimportant and a distraction from the ``real'' material.  To develop instructional materials and  methodologies that are thoughtful and engaging,  we must strive for balance: between texts and coding, between critique and solution, and between cutting-edge research and practical applicability. Finding such balance is particularly difficult in the nascent field of responsible data science (RDS), where we are only starting to understand how to interface between the intrinsically different methodologies of engineering and social sciences.   

In this paper we recount a recent experience in developing and teaching an RDS course to graduate and advanced undergraduate students in data science.    We then dive into an area that is critically important to RDS --- transparency and interpretability of machine-assisted decision-making, and tie this area to the needs of emerging RDS curricula. Recounting our own experience, and leveraging literature on pedagogical methods in data science and beyond, we propose the notion of an ``object-to-interpret-with''.   We link this notion to ``nutritional labels'' --- a family of interpretability tools that are gaining popularity in RDS research and practice.

With this work we aim to contribute to the nascent area of RDS education, and to inspire others in the community to come together to develop a deeper theoretical understanding of the pedagogical needs of RDS, and contribute concrete educational materials and methodologies that others can use.  All course  materials are publicly available at \url{https://dataresponsibly.github.io/courses}.

\end{abstract}

%% file: intro.tex
\section{Introduction}
\label{sec:intro}

As an emerging discipline, {\em responsible data science (RDS)} has yet to be codified as a course of study at most university campuses. Despite increasing demand, there is a lack of curricular materials that are available for adoption into university graduate programs, undergraduate majors, general education tracks, and interdisciplinary minors. The challenge is compounded by a shortage of faculty with the expertise to develop and teach RDS courses.

Although numerous ethics courses are available, with many focusing specifically on technology and computer ethics, pedagogical approaches employed in these courses rely exclusively on texts rather than on software development or data analysis.  For this reason, technical students often consider these courses unimportant and a distraction from the ``real'' material.  To develop materials and instructional methodologies that are thoughtful and engaging, and that help students gain knowledge and skills useful in their future careers, we must {\em strive for balance}: between texts and coding, between critique and solution, and between cutting-edge research and practical applicability. While finding such balance is both necessary and difficult in any educational effort, it is particularly difficult in the nascent field of RDS, where we are only just starting to understand how to interface between the intrinsically different methodologies of engineering and social sciences.   

In this paper we recount a recent experience in developing and teaching an RDS course to graduate and advanced undergraduate data science students at New York University.  In our description, we give specifics about both the content and the instructional style, with the hope that the course will be useful to others who are developing and teaching on this topic.   To further illustrate the challenges of teaching RDS, and to propose a path forward, we dive into an area that is critically important to RDS --- transparency and interpretability.  Specifically, we look at interpretability of the underlying computational elements of machine-assisted decision-making, and tie those to the needs of emerging RDS curricula. Recounting our own experience, and leveraging literature on pedagogical methods in data science and beyond, we propose the notion of an ``object-to-interpret-with'', which takes its inspiration from an important concept that emerged through constructivist practices ---
objects-to-think-with.   We link this notion to ``nutritional labels'' --- a family of interpretbility tools that are gaining popularity in RDS research and practice, and illustrate their use in project-based learning experiences.  

Our paper makes the following contributions:
\begin{itemize}
    \item We are among the first to look at the pedagogical implications of responsible data science, creating explicit parallels between cutting edge data science research, and cutting edge educational research.
    \item We developed and are teaching a unique course on responsible data science, geared primarily toward technical students.  We give a detailed description of our course here. 
    \item Focusing on transparency and interpretability, we propose best practices and concrete implementable techniques for teaching this important topic, for others to use.   Our proposal is based on our own teaching experience, and on insights from constructivist and inquiry-based educational practices. 
\end{itemize}

With this work we aim to contribute to the nascent area of RDS education.  We hope to inspire others in the community to come together to form a deeper theoretical understanding of the pedagogical needs of responsible data science, and to develop and share the much-needed concrete educational materials and methodologies, striking the right balance between research and practice.

%% file: experience.tex
\section{Our Experience in Teaching RDS}
\label{sec:experience}

We seek to connect emerging responsible data science (RDS) pedagogical practices with established pedagogical practices. Current theories in the learning sciences emphasize successful learning as active, collaborative, socially-constructed and situated.  Activities should incorporate these tenets. 
There is now a representative group of technical data science programs at the graduate level, and growing offerings at the undergraduate level~\cite{aasheim_data_2015, anderson_undergraduate_2014, de_veaux_curriculum_2017}. Institutions such as Duke, Harvard, Oxford, Stanford, University of Michigan,  University of Texas, and others have recently introduced ethical data science courses, all of which approach the topic from humanistic or social science-based --- cultural, legal, and/or philosophical --- approaches.  What is less represented and what the course that we introduce explores is ethical issues from the point of view of the technical.  

We now recount a recent experience in designing and teaching an RDS course to graduate and advanced undergraduate students in data science at New York University. This  technical course tackles the issues of ethics, legal compliance, data quality, algorithmic fairness and diversity, transparency of data and algorithms, privacy, and data protection. The course leverages established learning theories and best pedagogical practices, detailed in Sections~\ref{sec:objects} and~\ref{sec:methods}.

\subsection{Course Design}
\label{sec:experience:design}

The RDS course is a semester-long course, structured as a sequence of lectures, with supplementary readings, labs, and accompanying assignments.  All course  materials, including the syllabus, weekly reading assignments, complete lecture slides, and lab assignments, are publicly available on the course website at \url{https://dataresponsibly.github.io/courses}.  Homework assignments, with solutions and grading rubrics, and a detailed description of the course project, will be made available to instructors upon request.

The RDS course has Introduction to Data Science, or Introduction to Computer Science, or a similar course as its only a prerequisite.  Machine Learning is not a prerequisite for the course.  This is a deliberate choice that reflects our goals to (1) educate data science students on ethics and responsibility early in their program of study, and (2) to enroll a diverse group of students.  Students are expected to have basic familiarity with the python programming language, which is used in labs and  assignments.

One of the challenges we faced when designing this course was the lack of a text book that offers comprehensive coverage of responsible data science, balancing case studies, fundamental concepts and methodologies from the social sciences, and statistical and algorithmic techniques.  As a result, the course does not have a required textbook. Each topic is accompanied by required reading. In some cases, expert-level technical research papers are listed as assigned reading.  However, important concepts from the assigned paper are covered in class, and students are instructed on where to focus their attention while reading the papers, and which parts to skim or skip. 

The course comprises six thematic modules.  Each module is presented using a combination of case studies, often from the recent press, fundamental algorithmic and statistical insights, and hands-on exercises using open-source datasets and software libraries. 
 
\paragraph{Module 1: Introduction and background (1 week)}  Course outline, aspects of responsibility in data science through recent examples.  We now describe this topic in some depth, to give the flavor of the instruction. (Other modules are described here in somewhat less depth.) Fairness and diversity are motivated by ``the classics'': the 2012 Wall Street Journal piece on Staples online price discrimination~\cite{staples}, the 2015 AdFisher study of gender and job ads~\cite{DBLP:journals/popets/DattaTD15}, and ProPublica's 2016 ``Machine Bias''~\cite{Angwin2016}.  Following a brief description of these case studies, we formalize fairness in classification, and start fixing the necessary terminology: statistical vs. structural bias, outcomes and populations, statistical parity.  We then give a preview of the fairness impossibility results, to immediately dispel any belief that technology alone can mitigate fairness and diversity issues.  Transparency and accountability are introduced using the 2012 racially identifying names in online ad delivery study by Latanya Sweeney~\cite{DBLP:journals/queue/Sweeney13}.  This study is used to start diving into the complexity and the opacity of the online ad delivery ecosystem, discussing which stakeholders may be responsible for racial bias, and which should be held accountable.  Data protection is motivated by the 1979 Barrow, Alaska alcohol study, highlighting that privacy is important, but is not the only concern.  We conclude the introduction with an overview of the data science lifecycle.

{\em Lab 1} reproduces a portion of the analysis from ProPublica's ``Machine Bias'' investigation~\cite{Angwin2016}.

\paragraph{Module 2: The data science lifecycle (2 weeks)} Overview of the data science lifecycle. Data profiling and validation.  Data cleaning.  Building an algorithmic foundation for data profiling~\cite{DBLP:journals/vldb/AbedjanGN15} and data cleaning~\cite{DBLP:journals/pvldb/ChuI16} tasks using the relational model, frequent itemset and association rule mining. Making a link between data quality and representativeness, and fairness of the machine learning methods trained on that data, using the analysis from ``To Predict and Serve''~\cite{lum}.

{\em Labs 2 and 3} give students some hands-on experience with data profiling capabilities of standard python libraries, and introduce them to a  state-of-the-art data imputation library called Datawig~\cite{Biessmann2018}.

\paragraph{Module 3: Algorithmic fairness and diversity (3 weeks)}  A taxonomy of fairness definitions; individual and group fairness~\cite{DBLP:conf/innovations/DworkHPRZ12,DBLP:journals/corr/FriedlerSV16,mitchell}. The importance of a socio-technical perspective: stakeholders and trade-offs. Impossibility results~\cite{DBLP:journals/corr/Chouldechova17,DBLP:conf/innovations/KleinbergMR17}; causal definitions~\cite{DBLP:conf/sigmod/SalimiRHS19}; fairness beyond classification and risk assessment~\cite{DBLP:conf/ssdbm/YangS17}.  Diversity in data science pipelines.  The Rooney rule, diversity in hiring and  admissions~\cite{DBLP:conf/edbt/StoyanovichYJ18}.  The technical meat of this module draws from recent algorithmic methods for fairness and diversity in classification, selection, and ranking, and ties back these technical approaches to philosophical and legal doctirnes, and regulatory requirements. 

{\em Labs 4, 5 and 6} continue the exploration of ProPublica's ``Machine Bias'' investigation, and introduce students to IBM's open source AIF360 toolkit.

\paragraph{Module 4: Transparency and interpretability (2 weeks)} Reasons for transparency and interpretability: human-in-the-loop, responsibility, trust.  Auditing black-box models; explainable machine learning.  Methods to generate local explanations.~\cite{ribeiro_why_2016}  Causal influence of features with Quantitative Input Influence~\cite{datta_algorithmic_2016}.  Discrimination in online ad delivery.  The AdFisher study revisited: stakeholders, responsibility, and legal ramifications~\cite{DBLP:conf/fat/DattaDMMT18}.  Ongoing efforts to regulate online ad delivery, due to potential for discrimination in housing and employment. 

{\em Labs 7 and 8} introduce students to Locally Interpretable Model Explanations (LIME)~\cite{ribeiro_why_2016} and to  nutritional labels for rankings ~\cite{DBLP:conf/sigmod/YangSAHJM18}.  

\paragraph{Module 5: Privacy and data protection (2 weeks)}  Overview of responsible data sharing. Anonymization techniques and the limits of anonymization. Harms beyond re-identification. Differential privacy.   Technical material includes randomized response, the fundamental theorem of data reconstruction~\cite{DBLP:conf/pods/DinurN03}, and an introduction to differential privacy~\cite{DBLP:journals/cacm/Dwork11}, with a discussion of the formal definition, the trade-off between privacy and utility, query sensitivity,  query composition, and privacy-preserving synthetic data generation.  A discussion of the Netflix de-identification attack, and of the ongoing debate about the use of differential privacy in the US Decennial Census~\cite{mervis_can_2019}.

{\em Lab 7} introduce students to secure hashing.  {\em Lab 8} gives students hands-on experience with privacy-preserving synthetic data generation, using an open-source toolkit~\cite{DBLP:conf/ssdbm/PingSH17}.

\paragraph{Module 6: Legal frameworks, codes of ethics, professional responsibility (2 weeks)}  Ethical principles and frameworks (using~\cite{salganik} Chapter 6 as reading).  The Belmont Report and the Menlo Report.  The ACM Code of Ethics and Professional Conduct.  Legal frameworks: overview of thhe General Data Protection Regulation, and of the emerging algorithmic transparency regulation in the US.    

\paragraph{Homework assignments} The course includes four homework assignments, completed individually by the students.  A homework corresponds to approximately 2 weeks of effort.  All labs and most homework assignments are formulated as Jupyter notebooks, and labs serve as a starting point for completing the homeworks. While none of the homeworks are stand-alone writing assignments, writing is embedded in the homeworks. During the homeworks, students write computer programs and provide short essays on the interpretation of the data, and the implications around some decision or problem that the data informs.  The course also includes one problem-based homework that focuses primarily on (breaking) anonymization and on differential privacy.   

Developing homework assignments, particularly problem-based ones, constituted a significant effort during course design.  For example, we were surprised to find that virtually no problem sets are available for differential privacy --- a topic that is perhaps the best established among those we covered, and one that is starting to see broad practical adoption.  While there are some teaching materials on differential privacy for advanced graduate students in computer science, we were unable to find any existing materials that would be suitable for beginning data science students, and for data science practitioners.      

\paragraph{Course project} The project is completed in teams of two students, and corresponds to approximately 4 weeks of effort. The project is customized per course offering, and we describe it in more detail in the following section.

\subsection{Teaching and Iterating}
\label{sec:experience:teaching}

The RDS course was offered to data science graduate students (and advanced undegraduates) for the first time in Spring 2019, and will be offered at least once annually going forward.  Additionally, we are developing an undergraduate RDS course --- a program requirement of the newly-estabished data science major at the Center for Data Science at NYU, with plans to offer it at least once annually as well, starting in Spring 2021.  In addition to stand-alone courses, we are incorporating RDS modules into existing computer science and data science curricula, with materials based largely on those we developed for the RDS course described here, and those we will develop in the future.  

\paragraph{Student diversity and differentiated instruction} One of the most rewarding, but also the most challenging, aspects of teaching RDS was the diversity of the students. Students come into data science from different academic backgrounds, including computer science, mathematics, statistics, the natural sciences, social science, and law.  This full diversity of backgrounds was represented in our first student cohort, with students bringing their interdiscipinary perspectives to discussions.  On the other hand, this also meant that students were of different levels of technical preparation --- some had more of a familiarity with algorithmic techniques, and came in with more substantial programming and data analysis experience then others did.

One focus of the next iteration will be to explore purposeful differentiated instruction, a pedagogical approach that tailors instruction (e.g., content, process, environment) to address variance in students’ prior knowledge, disciplinary knowledge, learning styles, accessibility needs, and other background characteristics~\cite{chamberlin_promise_2010,george_rationale_2005,tomlinson_how_2017}.  

Differentiated instruction is an inclusive approach that does not equate to diluting a course for the less experienced nor does it equate to gearing course content towards the most advanced. Rather, it involves providing flexible learning environments so that all students progress towards learning goals in a way that recognizes the diversity of the group.  For example, a course with technical experts and technical novices can entail a group project where novices must take on a coding role, with advanced students serving as project leader and providing guidance to those with less experience.  Students with specific disciplinary backgrounds can peer review contributions made by students with little experience.  Essential to differentiated instruction is ongoing (formative) assessment so that the course dynamically meets the needs of students.  

\paragraph{Course project}
During the first offering of the course, students were directed to develop an interpretability tool for an automated decision system (ADS) of their choice, based on the concept of a nutritional label.  We suggested that students develop a nutritional label for one of the systems developed in response to a Kaggle competition of their choice, but students were encouraged to look beyond Kaggle.  Project deliverables included a written report, an implementation of the nutritional label in python, and a project presentation.   To encourage students to critically evaluate the system they are describing, we specified the following structure for the written report:

\begin{enumerate}
 \item {\bf Background}: general information about your chosen ADS. (a) What is the purpose of this ADS?  (b) What are its stated goals? (c) If the ADS has multiple goals, explain any trade-offs that these goals may introduce.
 
\item {\bf Input and output}. (a) Describe the data used by this ADS. How was this data collected or selected?   
(b) For each input feature, describe its datatype, give information on missing values and on the value distribution.  Show pairwise correlations between features if appropriate. (c) What is the output of the system, and how do we interpret it?

\item {\bf Implementation and validation}: present your understanding of the code that implements the ADS.  This code was implemented by others (as part of the Kaggle competition), not by you as part of this assignment.  Your goal here is to demonstrate that you understand the implementation at a high level. (a) Describe data cleaning and any other pre-processing. (b) Give high-level information about the implementation of the system. (c) How was the ADS validated?  How do we know that it meets its stated goal(s)? 

\item {\bf Outcomes}. (a) Analyze the effectiveness (accuracy) of the ADS by comparing its performance across different sub-populations.  (b) Select one or several fairness or diversity measures, justify your choice of these measures for the ADS in question, and quantify the fairness or diversity of this ADS.  (c) Develop additional methods for analyzing ADS performance: think about stability, robustness, performance on difficult or otherwise important examples, or any other property that you believe is important for this ADS. 

\item {\bf Summary}. (a) Do you believe that the data was appropriate for this ADS? (b) Do you believe the implementation is robust, accurate and fair / diverse / stable ? (c) Would you be comfortable deploying this ADS in the public sector, or in the industry?  Why so or why not?  (d) What improvements do you recommend to the data collection or analysis methodology?  
\end{enumerate}

All student teams delivered insightful analysis as part of their project submissions, demonstrating good command of material on algorithmic fairness, diversity, transparency, and interpretability.

\paragraph{Student feedback} The course was well-received by the students, who commented favorably both on the content, and on the pace.  Students felt that the course was relevant to their data science goals and career objectives, and that programming assignments were consistent with the content of the lectures.  
Free-text comments included primarily praise, along with critical feedback. Most notably, several students requested to expand the coverage of transparency and explainable models.

One of the students summarized his experience as follows:\newline 

\indent{\em In both my academic education and professional experience, I always sensed the need to make a deep dive on the topics we discussed in class. I've never really felt ``at ease'' when working with data to making decisions that impact other people’s lives, [...].  I tried many times in my early career to explain that deciding whether to use the mean or the median (or other quantiles) for setting an insurance rate is an ethical decision. Unfortunately, many believe that data and algorithms are the fairer way to assess things and drive decision because of their alleged objectivity. In reality, I think that many hide behind data, statistics, and science as a ``safe'' way to avoid the hard work to take an ethical stance.

Your class gave me structure, knowledge, and tools on how to go about these topics more credibility when discussing these issues with others and put decisions into actions. I am now thinking about ways to expand from what I learned in class, and eventually to put together a set of demos to better narrate these topics to non-technical audiences. I will keep you posted on how this goes!} \newline 

Several students suggested to expand on algorithmic transparency because {\em ``... it's a really interesting topic that is useful for both fairness and actually building better models.''} The next iteration will include targeted assessments that gauge students’ prior knowledge, suitability for group work, and improved mapping to other courses that feed into the course or would be taken afterwards. In the remainder of this paper, we focus on transparency and interpretability, giving some socio-technical background (in Section~\ref{sec:interpret}), and then treating these topics through the lens of data science education (in Sections~\ref{sec:objects} and~\ref{sec:methods}).

%% file: interpret.tex
\section{Transparency and Interpretability}
\label{sec:interpret}

An essential ingredient of successful machine-assisted decision-making, particularly in high-stakes decisions, is \emph{interpretability} --- allowing humans to understand, trust and, if necessary, contest, the computational process and its outcomes.   These decision-making processes are typically complex:  carried out in multiple steps, employing models with many hidden assumptions, and relying on datasets that are often repurposed --- used outside of the original context for which they were intended.\footnote{See Section 1.4 of Salganik's ``Bit by Bit''~\cite{salganik} for a discussion of data repurposing in the Digital Age, which he aptly describes as ``mixing readymades with custommades.''}  In response, humans need to be able to determine the ``fitness for use'' of a given model or dataset, and to assess the methodology that was used to produce it.

In the remainder of this section, we review the state of the art in transparency and interpretability, providing a  foundation for these concepts from the point of view of the discipline of responsible data science (RDS).  This foundation is necessary to inform Section~\ref{sec:methods}, where we propose a methodology for teaching these topics. This foundation also highlights the need for a balance between the technical consumption of machine learning models, built on accuracy and other performance metrics, and the contextual understanding of models, built on an interdisciplinary understanding of how the model behaves.

In our discussion, we make a distinction between {\em model transparency} and {\em data transparency} that somewhat mirrors the data science lifecycle.  Much of current research on transparency and interpretability in machine learning has focused on the last mile of data analysis --- on model training and deployment.  The discourse on model transparency, discussed in Section~\ref{sec:interpret:models}, pertains primarily to these lifecycle stages.  Several lines of recent work argue that critical opportunities for improving data quality and representativeness,  controlling for bias, and allowing humans to oversee and influence the process are missed if we do not consider earlier lifecyle stages~\cite{Kirkpatrick:2017:AD:3042068.3022181,LehrOhm2017,DBLP:conf/ssdbm/StoyanovichHAMS17}.  In particular, Lehr and Ohm~\cite{LehrOhm2017} underscore the need to interrogate data selection, collection, cleaning, and other kinds of preprocessing --- stages to which they refer as ``Playing with data.''  The discourse on data transparency, described in Section~\ref{sec:interpret:data}, pertains to these lifecycle stages.  We reconcile the notions of model and data transparency in Section~\ref{sec:interpret:labels}, where we discuss a unifying approach based on the concept of a nutritional label.

\subsection{Model Transparency and Interpretability}
\label{sec:interpret:models}

Transparency and interpretability are central to the critical study of the underlying computational elements of machine learning platforms, and can allow for a host of addressable questions~\cite{diakopoulos_accountability_2016, goodfellow_explaining_2015}.  Keeping dataset creation and preprocessing decisions in mind, how is data transformed once accessible to the platform? Which features of the input  most affect the algorithm’s outputs?  Who has access to the underlying model?  Can stakeholders understand what a model means?  Scholars from humanistic, social science, and scientific backgrounds voice the importance of introducing ethical and human-centered approaches when thinking about transparency and interpretability~\cite{gilpin_explaining_2018, goodman_european_2017, lage_human---loop_2018, leonelli_locating_2016, markham_ethics_2018}. %If so, which stakeholders and how?

Importantly, several scholars note a long history of conflating interpretability and explainability, and the distinction is important since “measurable data [associated with interpretability and prediction] are not accurate representations of their underlying constructs [associated with explainability]”~\cite[p. 293]{shmueli_explain_2010}.  Interpretability refers to the ability in which a cause and effect can be observed within a system, or how much one is able to predict what is going to happen, particularly in the context of specific inputs or parameters~\cite{carmichael_data_2018, doshi-velez_towards_2017, guidotti_survey_2018, DBLP:conf/kdd/LouCG12}.  One can clearly intuit what the model is doing by seeing a clear and consistent relationship between the set of features and model specifications, and the resulting metric.  In contrast, explainability is the extent to which the internal mechanics of an algorithmic system can be communicated to people.  Of course we have a host of audiences implied which complicates explainability - a model may be explainable, but to whom?  

For a system to be interpretable, one must comprehend what the model did in that context, and is primarily an empirical exercise in gauging the appropriateness of the predictive model and the resulting prediction.  One can comprehend a predictive metric and intuit the overall mechanics of the system without understanding why the system has produced the metric.  For a system to be explainable, a person has to explain what is happening and this done at the construct level. For Gilpin and colleagues, “interpretability alone is insufficient. In order for humans to trust black-box methods, we need explainability – models that are able to summarize the reasons for neural network behavior, gain the trust of users, or produce insights about the causes of their decisions. While interpretability is a substantial first step, these mechanisms need to also be complete, with the capacity to defend their actions, provide relevant responses to questions, and be audited … Explainable models are interpretable by default, but the reverse is not always true~\cite [p. 1]{gilpin_explaining_2018}. 

To achieve causal and other scientific explainability, an understanding of the holistic set of interpretations, formalisms, and predictions is required (Shmueli, 2010, Sweeney, 2017).  Lipton (2017) introduces granular and holistic variations in his notion of interpretability, which touch on the subtle differences between explainability and interpretability, even if: 

\begin{enumerate}
\item Simulatability occurs when one can understand the model as a whole.  A simple decision tree is understandable since it ``can be readily presented to the user with visual or textual artifacts''~\cite{ribeiro_why_2016}. This model is in contrast with, for example, a complex neural network model, where there are many implicit rules and dependencies, making it difficult to achieve a holistic interpretation of the model. 
\item Decomposability happens when ``each input, parameter, and calculation [of a model] admits an intuitive explanation''~\cite[p. 5]{lipton_mythos_2016}. Specifying the audience and purpose of interpretability is essential, as is articulating the desired balance between model performance and model interpretability~\cite{breiman_statistical_2001, gleicher_framework_2016}.
\end{enumerate}

Achieving both lines of interpretability can be important to address trust, legality, and other desirable aspects ~\cite{luban_moral_1992}.  In addition to Lipton~\cite{lipton_mythos_2016}, other scholars have highlighted the lack of a universal definition and purpose of interpretability in the context of machine learning~\cite{carmichael_data_2018, doshi-velez_towards_2017, DBLP:conf/kdd/LouCG12}.  We take interpretability to generally mean a person’s or group’s ability to understand a model --- for example, to describe what the inputs are, how the algorithms operate, how outputs are framed, and even how to articulate the model’s reuse over time and contexts.  Specifying the audience and purpose of interpretability is essential, as is articulating the desired balance between model performance and model interpretability~\cite{breiman_statistical_2001,gleicher_framework_2016}.  For Guidotti et al.~\cite{guidotti_survey_2018}, an interpretable model is one that demonstrates accuracy and fidelity, as well as allows for human interpretability. 

Selbst and Barocas~\cite{selbst_intuitive_2018} add to the interpretability/explainability debate, incorporating legal, technical, and philosophical discussions.  If linearity, monotoneity, continuity, and dimensionality are the four qualities that drive model complexity, then the concepts of inscrutability and non-intuitiveness are the two primary interpretability focal points upon which we should concentrate.  For the authors, inscrutability equates to Lipton’s definition of simulatability; unintuitiveness refers to a lack of instinctual understanding of the underlying statistical relationships at play, even if those relationships have been explicitly revealed.  

We take interpretability to generally mean a person’s or group’s ability to understand a model -- for example, to describe what the inputs and algorithms are, how the algorithms operate, how outputs are framed, and even how to articulate the model’s reuse over time and contexts.  Specifying the audience and purpose of interpretability is essential, as is articulating the desired balance between model performance and model interpretability (Breiman 2001; Gleicher 2016).

Another consideration in achieving interpretability is the intended audience.  The most substantive publications around  transparency and interpretability are geared towards technical audiences and offer varying levels of accessibility for non-technical groups~\cite{ananny_seeing_2016,datta_algorithmic_2016,DBLP:journals/cacm/Dwork11,kim_scalable_2015,lage_human---loop_2018,ribeiro_why_2016}. Publications geared towards the wider public generally fall under data journalism~\cite{broussard_artificial_2018,mervis_can_2019,oneil_weapons_2016}.  Gillborn, Warmington and Demack~\cite{gillborn_quantcrit:_2017}, in the context of critical data studies, highlight the hidden assumptions that characterize algorithms, platforms, and intended audiences.  Establishing interpretable model insights for a technical audience is fairly straightforward, but there is a host of challenges to consider when aiming for interpretability for lay audiences, including audience expectations, relevance of the algorithm or platform, and audience preparedness.  Based on articles that call for machine learning interpretability, authors assume that the public will:
\begin{itemize}
\item expect relevant and useful data to be made available;
\item have the time, resources, and expertise to access the data and then analyze it; and
\item will alter behavior if the data reveals poor equality results.
\end{itemize}

These assumptions, while noble, are often presumptuous, and can lead to specific design decisions that may in turn not lead to widespread understanding of complex concepts.  For simplicity, and based our teaching experience, we focus on a population whose technical sophistication and interest in real-world problem solving lies in the middle --- students of data science.  Targeting this audience connects to established ways in which people learn about a complex topic --- what they are able to find interpretable --- while pinpointing generalizable aspects that hold potential for the wider public.

\subsection{Data Transparency and Interpretability}
\label{sec:interpret:data}

In a recent essay, Stoyanovich and Howe argue that data transparency is an essential component of algorithmic transparency~\cite{follow}.  We recount these arguments here.

In applications involving predictive analytics, data is used to customize generic algorithms for specific situations --- algorithms are trained using data.  The same algorithm may exhibit radically different behavior --- make different predictions; make a different number of mistakes, and even different kinds of mistakes --- when trained on two different datasets. In other words, without access to the training data, it is impossible to know how an algorithm would actually behave. 

Algorithms and corresponding training data are used, for example, in predictive policing applications to target areas or people that are deemed to be high-risk.  But as has been shown extensively, when the data used to train these algorithms reflects the systemic historical bias towards poor and predominately African American neighborhoods, the predictions will simply reinforce the status quo rather than provide any new insight into crime patterns.  The transparency of the algorithm is insufficient to understand and counteract these particular errors.  Rather, the conditions under which the data was collected, the data processing methodology, and the resulting composition of the training dataset must be retained and made available to make the decision-making process transparent.

Even those decision-making applications that do not explicitly attempt to predict future behavior based on past behavior are still heavily influenced by the properties of the underlying data. For example, the VI-SPDAT~\cite{VISPDAT} risk assessment tool, used to prioritize homeless individuals for receiving services, does not involve machine learning, but still assigns a risk score based on survey responses --- a score that cannot be interpreted without understanding the conditions under which the data was collected.  As another example, matchmaking methods such as those used by the Department of Education to assign children to spots in public schools are designed and validated using datasets; if these datasets are not made available, the matchmaking method itself cannot be considered transparent.

Data transparency is important both when an automated decision system is interrogated for systematic bias and discrimination, and when it is asked to explain an algorithmic decision that affects an individual.  An immediate, and often impractical, interpretation of data transparency is making the training and validation datasets publicly available.  However, while data should be made open whenever possible, much of it is sensitive and cannot be shared directly. That is,  data transparency is in tension with the privacy of individuals who are included in the dataset.  In light of this, an alternative interpretation of data transparency is as follows: 

\begin{itemize}

\item In addition to releasing training and validation datasets whenever possible, summaries of relevant statistical properties of the sensitive datasets can be made available, to aid in interpreting the decisions made using the data, while applying state-of-the-art methods to preserve individuals' privacy.

\item When appropriate, privacy-preserving synthetic datasets can be released in lieu of real datasets to expose certain features of the data, if real datasets are sensitive and cannot be released to the public.

\end{itemize}

An example of a data transparency approach is the Datasheets for Datasets project~\cite{Gebru2018} that advocates for a  standardized process for documenting datasets.   Specifically, the authors propose that every dataset be accompanied with a datasheet that documents its motivation, composition, collection process, and recommended uses. The goal is to facilitate better communication between dataset creators and dataset consumers, and to encourage transparency and accountability in data use.  Notably, the database and cyberinfrastructure communities have been studying systems and standards for metadata, provenance, and transparency for decades~\cite{op,DBLP:journals/concurrency/MoreauLA08}.  These concepts are now seeing renewed interest in the context of transparency and interpretability.

\subsection{Nutritional Labels for Data and Models}
\label{sec:interpret:labels}

A novel approach to interpretability is the nutritional label. The most famous nutritional label, the Nutrition Facts panel, began appearing on all packaged foods after the passage of the Nutrition Labeling and Education Act of 1990~\cite{food_and_drug_administration_nutritional_1994}.  This panel and its antecedents evolved over time. Initial versions had the purpose of protecting the public from deceptive and dangerous information about food products, while the current version is geared towards empowering the public to make informed decisions over their nutritional habits.

Appropriating the nutritional label paradigm for machine learning is a logical one, both for communication purposes and for learning purposes.  For engaging the general public, it appropriates a familiar visual artifact to communicate highly technical and opaque information.  For facilitating the learning of data science concepts, it requires students to actively and iteratively come to understand hidden algorithms, distill the most important information about a model, and adapt that knowledge for non-technical audiences.  Nutritional labels synthesize information about machine learning models into a visually compact format; as a result, they obscure the more complex aspects of a model in the service of visual economy.  The nutritional label is not the singular, correct way to communicate the model to diverse audiences since it does not make a model definitively interpretable.  The format does, however, combine textual explanations and graphic information, representing a best practice of dual learning theory appropriate for learners~\cite{national_research_council_how_2000}.  Research supports this pedagogical advantage, demonstrating that nutritional labels, particularly those that have interactive functionality, can increase one's understanding of a complex topic and lead to better decision making~\cite{byrd-bredbenner_inherent_2009,gunaratne_using_2017,kelley_standardizing_2010}. The nutritional model as a paradigm creates an entry point at which one can engage and start to question the interpretability of a model.  In analyzing or creating a label while working at the technical level of the model, one is faced with the ways in which creating a singular, understandable presentation that works in all cases and for all audiences is in fact impossible. There is a utility in creating an artifact that signals certain problematic aspects of the model, particularly for learners with more technical sophistication.  

Two sets of scholars have explored the use of the nutritional label in data science. Ranking Facts is an application that reveals in a user-friendly way stability, statistical parity, and diversity measures associated with ranking algorithms~\cite{DBLP:conf/sigmod/YangSAHJM18,label,refining}. The assumption is that items placed in top ranks are of a higher quality than items placed lower down the ranked list (\eg item in rank 1 is of higher quality than item in rank 150).  This simple schema implies a high level of interpretability, though Ranking Facts reveals that rankings may be highly sensitive to inputs and can hide disparate impacts on subsets of data. Revealing this information as a visual gives a deeper understanding of the underlying algorithm, and puts any particular set of rankings into greater context. As seen in Figure 1, Ranking Facts assists a lay audience in achieving simulatability~\cite{ribeiro_why_2016}, model intuitiveness, and a level of transparency around the topics of fairness, diversity, and stability.  The Fairness pane supports an intuitive understanding of whether or not the model exhibits statistical parity without requiring knowledge of the term's mathematical properties.  Likewise, the Diversity pane signals how well the model represents categories.  

\begin{figure}[t!]
\centering
\includegraphics[width=.44\textwidth]{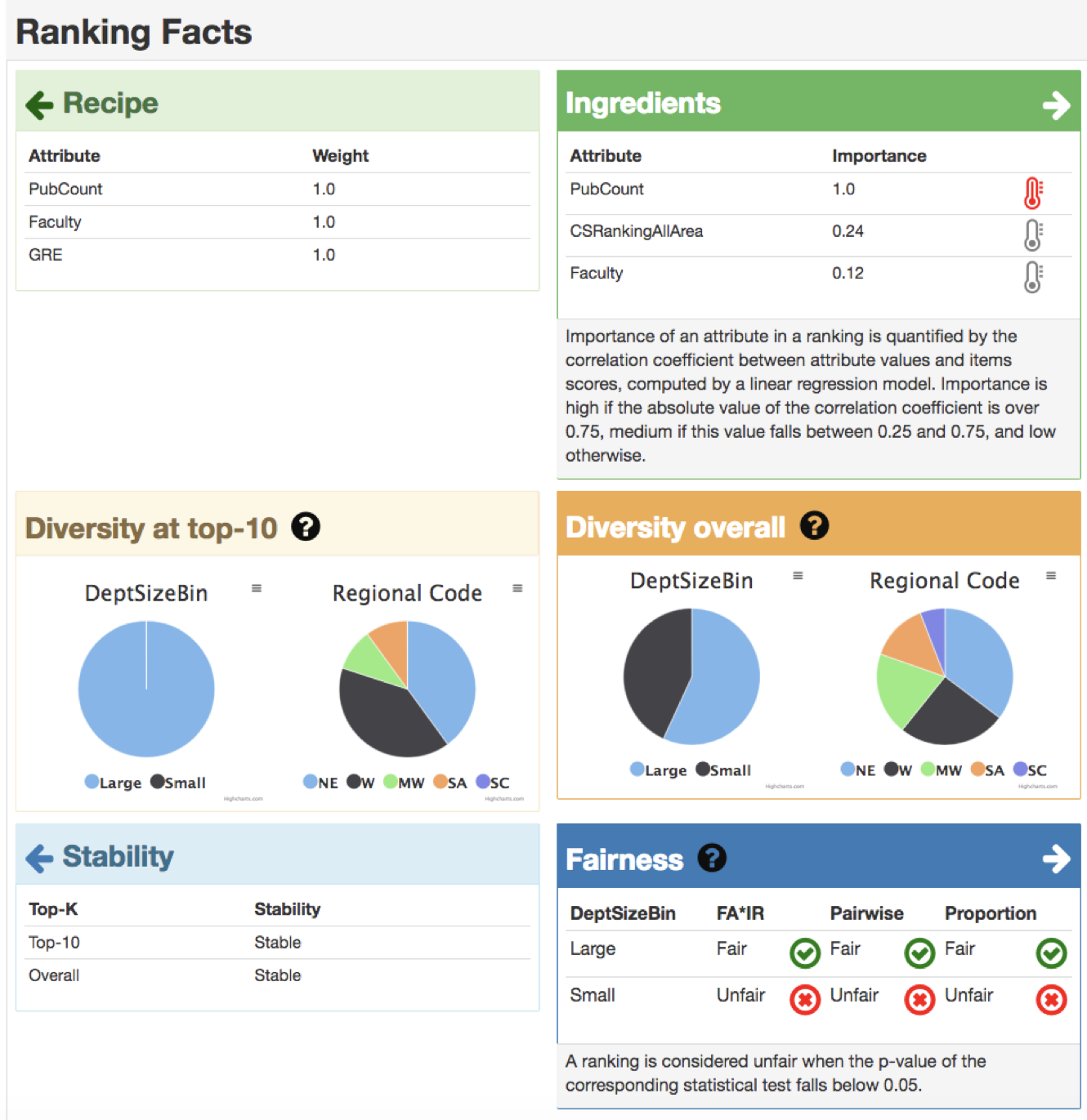}
\caption{The Ranking Facts nutritional label for rankings, from \url{http://demo.dataresponsibly.com/rankingfacts/}. This interpretable representation of a dataset of university department rankings is constructed automatically by the open-source web-based tool.}
\label{fig:ranking}
\end{figure}

\begin{figure}[t!]
\centering
\includegraphics[width=.44\textwidth]{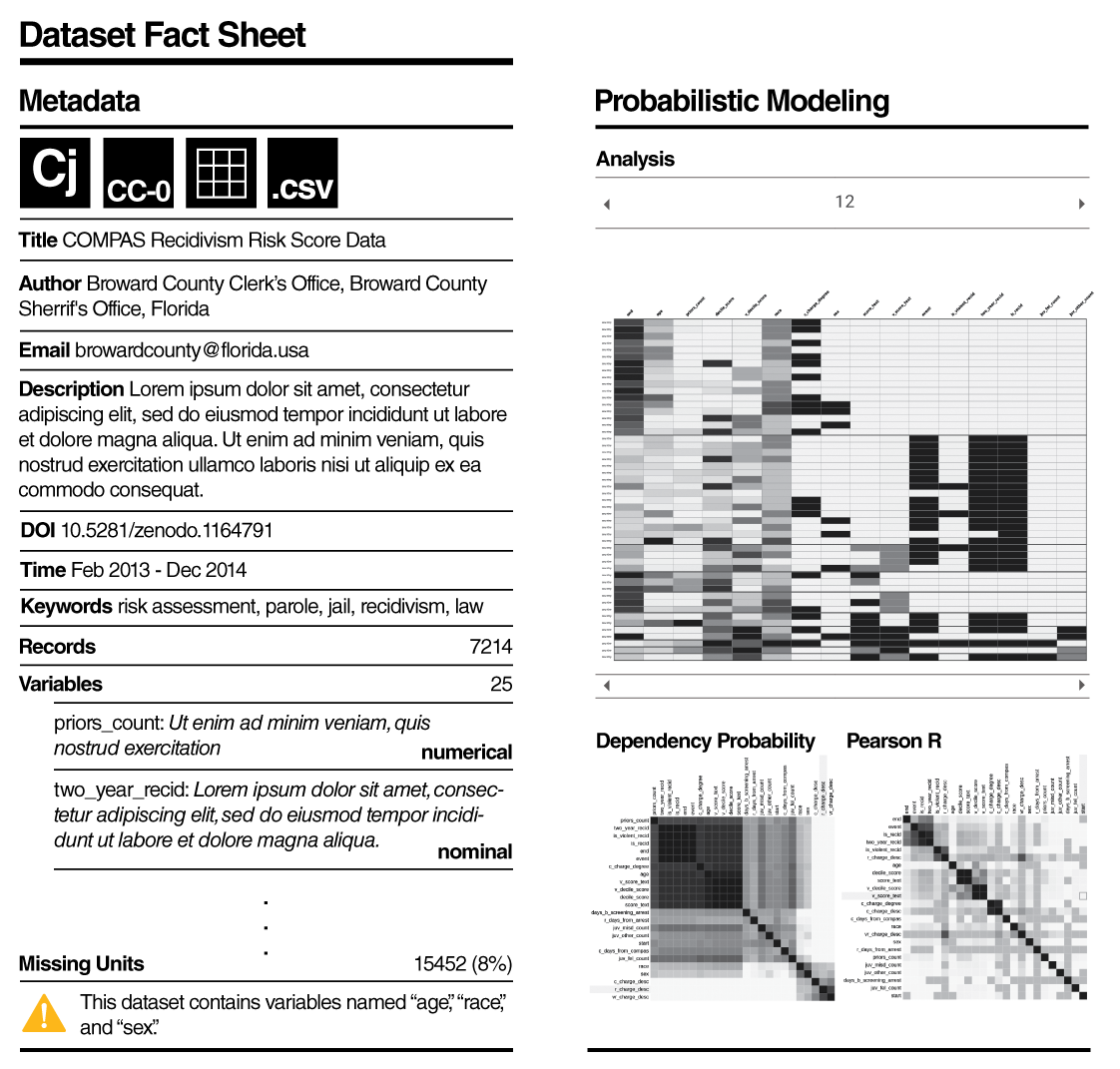}
\caption{The Dataset Fact Sheet prototype, from \url{https://ahmedhosny.github.io/datanutrition/}.  This representation of the COMPAS Recidivism Risk Score Dataset is manually constructed.}
\label{fig:datasetsheet}
\end{figure}

A second team~\cite{DBLP:journals/corr/abs-1805-03677} also used the nutritional label as an appropriate tool to think about algorithmic transparency.  The team developed a framework, the Dataset Nutrition Label, which provides modules that display metadata and source information, textual descriptions and summary statistics of variables, as well as graphs visualizing more complicated information like probabilistic models and ground truth correlations. The Metadata panel summarizes relevant dataset information while the Modeling pane provides model-specific information about performance and accuracy. Other panels go into detail about dataset authorship, model variables and ground truth correlations.  The Dataset Nutrition Label contributes largely to transparency by taking a descriptive snapshot of a dataset's inherent qualities (i.e. number of records and variables) and introducing some interpretable features (i.e. keywords).  The Probabilistic Modeling pane requires knowledge of a model's properties, though a highly visual presentation facilitates understanding.

%% file: objects.tex
\section{Data Science Education: Charting new pedagogical territory} 
\label{sec:objects}

There is a dearth of scholarship around data science education proper, since most formal academic programs have sprouted up only over the past 5 years or so.  Though data science education is still maturing, recent curricular guidelines have been developed~\cite{de_veaux_curriculum_2017} that emphasize theoretical and applied mathematical and computational knowledge. This is exemplified in a report from one of the rare long-term undergraduate data science programs.  The program, which includes computer science and mathematics courses, as well as three data science-specific courses: introductory data science, dataset organization and management, and a capstone course~\cite{anderson_undergraduate_2014}. A novel aspect to this program is the required completion of a cognate, which is a set of discipline-specific courses that support interdisciplinary applications of data science knowledge.  The guidelines also emphasize the integration of communication, reproducibility, and ethics into data science curricula.  “Programs in data science should feature exposure to and ethical training in areas such as citation and data ownership, security and sensitivity of data, consequences and privacy concerns of data analysis, and the professionalism of transparency and reproducibility”~\cite[p. 22]{anderson_undergraduate_2014}. RDS education is rare, particularly as a degree requirement.  

Scholars have noted that data science education blends statistics and computer science pedagogical principles, but that there remains a lack of integration between theoretical principles and real world applications~\cite{farahi_michigan_2018}.  Additionally, the interdisciplinary nature of data science necessitates a new kind of pedagogy, one that not only requires robust technical, theoretical, and practical STEM-based training, but also a humanist sensibility that highlights ethical concerns and communication challenges. We can, however, examine pedagogical scholarship that exists in computer science and statistics education to make logical claims about how students learn about data science, and develop best practices. 

\subsection{General Overview}
\label{sec:objects:general}

As in other STEM fields, constructivism and inquiry-based learning are the reigning pedagogical paradigms~\cite{disessa_knowledge_1988,national_research_council_how_2000,turkle_epistemological_1992}. Constructivism emphasizes the learning process as active, iterative, applied, and student-centered; similarly, inquiry-based learning makes the student an autonomous seeker of their own knowledge, involved as much in the question formation (what do I want to know?) as in the answer formation (what do I know and how?).  

Students learn computer science concepts through hands-on programming, connecting theory through applied activities and ongoing inquiry~\cite{ben-ari_constructivism_2001}.  Efforts also have been made to visualize, and to have students visualize, the mechanisms by which known algorithms operate~\cite{hundhausen_using_2000,sriadhi_rc4_2018}.  Empirical evidence from experimental design trials supports the fact that these graphics do support learning outcomes, but only if they maximize learner engagement (\ie they allow for interactivity like algorithm building)~\cite{hundhausen_meta-study_2002,naps_exploring_2002}. In statistics education, one focus has been on offering a holistic view of the data analysis process, compelling students to think about the creation of data as equally important as result significance~\cite{hilliam_interactive_2019}.  Another focus is on the importance of effectively-generated visualizations to communicate statistical concepts~\cite{garfield_how_2007,nolan_teaching_2016,tufte_visual_2001,tukey_exploratory_1977}. Variability rather than inference, for example, is a concept that is more readily understandable through visual representation~\cite{rubin_exploring_2006}.

The urgency of introducing ethical and social considerations into data science curricula cannot be overstated, particularly with current reports of the ways in which organizations have transgressed privacy and other laws with the help of data science.    Both the Association of Computing Machinery~\cite{association_for_computing_machinery_acm_2018} and the American Statistical Association~\cite{gaise_college_report_asa_revision_committee_guidelines_2016} have issued official guidelines with respect to ethics and ethical conduct for practitioners. There is established cross-disciplinary research on ethical reasoning~\cite{boyd_privacy_2010, knapp_engaging_2016, leonelli_locating_2016}, with specific recommendations for pedagogy and training that emphasize case studies, field practice, and sensemaking frameworks~\cite{mumford_sensemaking_2008, sternberg_teaching_2010}.  In their discussion on teaching ethical computing, Huff and Martin~\cite{huff_computing_1995} summarize a framework for introducing ethical analysis that incorporate levels of social analysis (individual through global), responsibility, privacy, reliability, equity, and other related topics.  In terms of pedagogy, they make an important recommendation of “incorporation of ethical and social issues in the lab work associated with such standard computer science subjects  as database design, human-computer interaction, operating systems, and algorithms”~\cite[p. 83]{huff_computing_1995} (83).  Tractenberg and colleagues~\cite{tractenberg_using_2015} offer guidelines on introducing ethical reasoning into data science training, and detail two syllabi that have students reflecting on ethical misconduct, societal impacts, privacy and confidentiality considerations, and responsible research practices, though evaluation hinges on written assignments and class discussion only.  There do exist courses that could be categorized as RDS courses, though they focus on ethical data science from a humanistic rather than technical perspective.  There also exist highly technical courses that touch on RDS topics, though not comprehensively~\cite{friedman_computing_1990, harvard_university_embedded_2019, martin_non-apologetic_1992, quinn_teaching_2006}.  The course that we discuss in this paper is unique in that it centralizes RDS topics in a highly technical data science course.

\subsection{Objects-to-interpret-with: Teaching for and Evaluating the Understanding of Data and Complex Models}
\label{sec:objects:obj}

Here we introduce our notion of students having \textit{objects-to-interpret-with}, which takes its inspiration from an important concept that emerges through constructivist practices --- objects-to-think-with~\cite{papert_mindstorms:_1980}. These objects are representations that help students grapple with universal concepts and ``understand how ideas get formed and transformed when expressed through different media, when actualized in particular contexts, when worked out by individual minds''~\cite{ackermann_piagets_2001}. Our objects of focus likewise assist students in forming heuristic knowledge and understanding contextual knowledge, but specifically target metacognition surrounding ways of interpretation and representation.

Objects-to-interpret-with also stem from Latour’s claim that artifacts that aid in interpretation do so by helping us ``understand how the mobilization and mustering of new resources is achieved''~\cite[p. 6]{latour_visualisation_1986}. These objects are never neutral, and elicits the user as a co-creator of knowledge~\cite{gitelman_introduction_2013}. The best public-facing systems adopt a prosumption view  that ``enables governing knowledge to appear as the product of co-creation rather than an expert technical and methodological accomplishment. It appears to normalize, neutralize and depoliticize statistical analysis''~\cite[p. 133]{williamson_digital_2016}. This more theoretical framing is important because it grounds the object-to-interpret-with as a tool to reveal the partiality and latent mechanisms of underlying algorithms, platforms that run such algorithms, and unforeseeable results that happen when certain datasets interact with machine learning algorithms and platforms.

Objects-to-interpret-with allow us to articulate exactly what students learning to interpret complex machine learning models should know.  Lipton~\cite{lipton_mythos_2016}, and Selbst and Barocas~\cite{selbst_intuitive_2018} detail the following technical ways to promote interpretability. Exposure to these maps to different types of knowledge in instructional objectives:
\begin{itemize}
\item Engineering algorithmic transparency so that the system reveals how sets of inputs lead to certain outputs, and that there is consistency regardless of the diversity in inputs.  For example, one can make explicit feature choices or parameters, or incorporate regularization; 
\item Developing posthoc methods for models that are very complex and/or remain opaque due to business necessity reasons.  These methods allow users to understand model outputs and potentially glean information about how different combinations of inputs yield different results without having low-level access to model specifics;
\item Creating interactive platforms that allow users to develop their own understanding about model functioning through consistent manipulation of parameters, inputs, and independent variables.
\end{itemize}
Interacting with systems engineered for algorithmic transparency supports the development of procedural and strategic knowledge, since it simplifies and makes explicit the process of transforming inputs into outputs, and supports the understanding of the model as general method.  Working with posthoc methods develops conceptual thinking, focusing on the model as a schema with multiple interpretations. Experimenting with interactive platforms develops metacognitive thinking and heuristic knowledge, where students implicitly build beliefs about their own learning.

For data science students, the focus is not necessarily on building algorithms, but rather on iterative question formulation and transforming vague goals into measurable parameters~\cite{passi_problem_2019}. Objects-to-interpret-with promote making sense of ill-defined information and indefinite meanings to achieve  deeper learning.  In order for interpretability to occur, a deeper than normal understanding of how inputs lead to outputs is required, and the aforementioned approaches provide more or less of a scaffolding, depending on the sophistication of the student.   Additionally, there is a collaborative, constructivist nature to learning~\cite{wenger_communities_1998}. When teaching complex machine learning models to data science students, particularly with a focus on fairness and transparency, it is beneficial to provide diverse presentations to facilitate the development of heuristics, accommodate diverse learning styles, and acknowledge the motley nature of the information being imparted. There are substantive interpretability learning opportunities found in examining isolated elements (\eg parameters, variables, nature of inputted data, and algorithms) and holistic models.  Lipton~\cite{lipton_mythos_2016} details the following modes of presentation that facilitate learning interpretation:
\begin{itemize}
\item Text explanations: best for metadata and overall contextualization.
\item Visualization: best for highlighting individual elements of a model.
\item Local explanations: best for black box models; Identifies ``an interpretable model over the interpretable representation that is locally faithful to the classifier''~\cite[p. 3]{ribeiro_why_2016}.
\item Example aggregation: best for metacognitive and holistic understanding of a model; allows for a learner to build a set of heuristics through exposure to diverse examples~\cite{lundberg_unified_2017,poursabzi-sangdeh_manipulating_2018}.
\end{itemize}

This list maps to pedagogical best practices that deliver learners information in ways that maximize understanding since it (1) presents information multimodally (visually, text-bases, and through case studies), (2) provides opportunities for active learning, and (3) develops metacognition.  Additionally, the presentation of complex machine learning models provide opportunities for explicit and implicit assessment.

%% file: methods.tex
\section{Data Science Education: Practical Considerations for Teaching RDS}
\label{sec:methods}

There remains a dearth of resources and pedagogical methodologies for data science education, particularly for teaching RDS. Section~\ref{sec:experience} details an RDS course that balances the need for students to engage with topics including data protection, fairness, and transparency from both a technical and  an interdisciplinary perspective. Course activities integrate collaborative and inquiry-based learning, allowing students to broadened their technical and domain knowledge by interacting with peers of varying expertise and backgrounds.

\subsection{Pedagogical Best Practices for Teaching Interpretability}
When thinking about teaching students about transparency and interpretability of data and models, it is important that activities should incorporate an understanding of how to optimize learning with a need to develop students' technical know-how and impart an ethical and contextual understanding.  In 2016, the Park City Mathematics Institute issued recommended guidelines for data science education~\cite{de_veaux_curriculum_2017}. In addition to promoting requisite theoretical and technical knowledge and skills, the group highlights as a principle, "Knowledge transference", which includes communication, as well as ethics and reproducibility. "Programs in data science should feature exposure to and ethical training in areas such as citation and data ownership, security and sensitivity of data, consequences and privacy concerns of data analysis, and the professionalism of transparency and reproducibility"~\cite[p. 2.8]{de_veaux_curriculum_2017}. No specific guidance is offered on how these terms may be defined and/or introduced in a data science context, and below we outline concrete ways to integrate these concepts with technical training, pedagogical best practices, and ethical grounding in mind.

Courses should naturally incorporate the standard data science process, including business and problem understanding, data preparation and munging, modeling, evaluating, and deployment~\cite{provost_data_2013}.  Passi and Barocas conducted an ethnographic study of how professional data scientists complete the iterative data science process in situ~\cite{passi_problem_2019}. In this study, they focus on problem formulation, ``as much an outcome of our data and methods as of our goals and objectives,'' as a metaphor for the larger data mining process that has constantly mutable and latent aspects, which must be reconciled to achieve some result. Students will develop a sense of the sometimes amorphous nature of doing data science in real-world situations.

We should build interdisciplinary frameworks for understanding transparency, interpretability, and other relevant concepts. Critical data studies reveals that platforms, as well as the inputs and algorithms of said platforms, are as much socially constructed as any other cultural artifact.  There is a growing recognition that, even though machine learning systems may have been created through business-minded, technology-focus perspectives, legal, philosophical, and socio-cultural critical perspectives are requisite considerations.  Emerging subfields like critical data studies, ethical artificial intelligence, and RDS represent this line of thinking. Infusing these perspectives into data science curricula should be a standard.

There is also a need to leverage into coursework real-world platforms that offer distinct definitions of transparency and interpretability.  Distinct versions of transparency and limits of interpretability are revealed through broad exposure to different types of platforms.

\begin{table*}[ht]
\caption{Pedagogical activities for teaching interpretability.}
\centering
\begin{tabular}{p{0.2\linewidth}p{0.35\linewidth}p{0.35\linewidth}}
\hline
\textbf{Activity} & \textbf{Description} & \textbf{Assessment method}\\
\hline
Replicate an existing study & Students reconstruct a (portion of a) published study and highlight any replication issues & \textit{Metrics evaluation}: Evaluation of quantitative scores.

\textit{Content analysis}: of replication issues discussion.\\
\hline
Reverse outline/engineer a platform & Students outline a machine learning platform to understand relationships between inputs and outputs. & \textit{Open coding}: A process stemming from HCI research in which students will explicitly label processes related to the platform.\\
\hline
Diagnostic learning logs & A meta activity where students outline points of understanding and confusion about machine learning concepts. & \textit{Points of confusion}: student outlines points of confusion at designated moments throughout a process.\\
\hline
Problem recognition tasks & Students are presented with a set of data science problems and an array of algorithmic transparency platforms and work to identify the best procedure to address the problem.  & \textit{Controlled experiment}: different groups are given different platforms with which to explore an identical question. Comparisons and debriefs reveal differences in transparency and interpretability.\\
\hline
Process analysis & Students reflect on their process of a deliverable in a meta-reflective exercise. & \textit{Cognitive walkthroughs}: form of self report where student outlines decisions made to produce label.\\
\hline
Peer review & Students evaluate other students' performance on an activity. & \textit{Surveys}: Quick method for assessing students' self perceptions about a task or prior knowledge.

\textit{Rubric creation}: Groups of students formulate the assessment parameters for their peers.\\
\hline
Before-After & Students iterate on a task, for example, tweaking a particular parameter or variable. & \textit{A/B Testing}: of model performance and interpretability.\\
\hline
Design and deploy & Students work in groups to deploy a system, potentially using the design sprint technique. & \textit{Rapid ethnography}: a technique that records behavior as students work together.\\
\hline
Create a nutritional label for a machine learning platform & This is a combination of a documented problem solution, where students’ understanding emerges implicitly through process-based explanations, and a focus on varying aspects. & \textit{Content analysis}: rubric based analysis of student-produced deliverable.~\cite{schraagen_introduction_2000}

\textit{Cognitive walkthroughs}: form of self report where student outlines decisions made to produce label.\\
\hline
\end{tabular}
\end{table*}

Activities should be offered through multiple presentation modes and levels of interactivity to maximize engagement and promote heuristics development~\cite{fayyad_information_2002}. Current learning science theories map these aspects as key to deeper learning that goes beyond the more superficial knowledge of novices (i.e., remembering, understanding, and applying) and approaches the more complex cognitive processes required of expert knowledge (i.e., evaluation, metacognition, and creation)~\cite{mayer_applying_2010,national_research_council_how_2000}.

We should offer students opportunities for documentation as explanation and proper evaluation.  “Careful validation … is not enough.  Normatively evaluating decision-making requires, at least, an understanding of: (1) the values and constraints that shape the conceptualization of the problem, (2) how these values and constraints inform the development of machine learning models and are ultimately reflected in them, and (3) how the output of models inform final decisions”~\cite[p. 1130]{selbst_intuitive_2018}.  There exists a range of activities that support metacognition, the act of thinking about one's thinking, that will assist students in thinking more holistically about how models perform and how we assess this performance.  Documenting acceptable metrics like F- and AUC scores as valid indicators of the technical performance of a model, but noting that these metrics do not necessarily assist in evaluating fairness and transparency.  Activities that compel students to contemplate their own thinking become rich opportunities to expose assumptions and knowledge gaps in the way that we evaluate the data science process.

Importantly, we should layer in assessment, both quantitatively and qualitatively, and both formatively and summatively~\cite{angelo_classroom_1993,lazar_research_2010}.  Assessing how well students achieve model performance and model interpretability is challenging, given the tension between the two goals.  The former is metrics-based and therefore quantitatively assessable, while the latter requires a mix of quantitative and qualitative methods to assess whether or not students can interpret a model or find it transparent.

\subsection{Pedagogical Activities for Teaching Interpretability}
\label{sec:methods:activity}

Pedagogical best practices should inform the development of learning goals, which in turn are used to develop specific activities.  Goals can focus on developing skills-building, problem formulation and solving, descriptive and procedural knowledge, heuristics, and more esoteric concepts like metacognition, cooperation, and creativity. Lang focuses on broad categories that lead to activities that support a range of students in building knowledge, understanding critically, and motivating their own learning. For learning within technical domains, these categories are supported through activities like worked examples, exposure to common and unusual problems, in-class group problem solving, explicit teaching of models, interaction with simulations, and reflection~\cite{aleven_effective_2002,baker_toward_2016,ben-ari_constructivism_2001,dweck_motivational_1986}.  If we focus more pointedly on learning RDS topics, we need additional pedagogical techniques.

Table 1 showcases several pedagogically-sound activities that work well for teaching students interpretability. These activities are suitable for diverse groups, for example, (1) those of varying technical and theoretical prior knowledge, (2) those from varied disciplinary backgrounds, and (3) those with different learning preferences.  These activities, in combination, also support the development of an individual student's knowledge transitioning from novice to expert level.

The RDS course described in Section~\ref{sec:experience} incorporates some of these described activities. The final course project (Section~\ref{sec:experience:teaching}) combines elements of a replication study, process analysis, design and deployment, and nutritional label design. Students replicate a model with an existing dataset and algorithm, and in the process identify transparency flaws, areas for improving interpretability, and ways to improve model performance. Future iterations of the course will tie in further pedagogical principles.

%% file: conc.tex
\section{Conclusion and Outlook}
\label{sec:conc}

In this paper we looked at the pedagogical implications of responsible data science, creating explicit parallels between cutting edge data science research, and cutting edge educational research.  We recounted a recent experience in developing and teaching a responsible data science course to graduate and advanced undergraduate data science students.  Further, focusing on transparency and interpretability, we proposed best practices and concrete implementable techniques for teaching this important topic, for others to use.  

We are excited to see the enthusiasm of students, data science practitioners, and instructors for responsible data science.  Given this enthusiasm, and the tangible need of both the industry and academia to welcome a new generation of responsible data scientists, we must come together as a community to meet the challenge of developing curricula and teaching responsible data science, while striking the right balance.
We are at the beginning of the road, and much work remains: in developing instructional methodologies and materials, creating assignments and assessment instruments, and ensuring that the materials we develop stay up-to-date as our understanding of ethics and responsibility in data science evolves.  We must also be deliberate in finding ways to scale up curriculum development and instructor training.

In this article we focused primarily on higher education, and in particular on teaching data science students.  Going forward, it is crucial to think about educating current data science practitioners, and members of the general public.  As with the data science student population, transparency and interpretability will prove to be a key concept to teach.  

A necessary next step is to advance the work of reconciling various disciplinary critiques of interpretability and explainability in machine learning~\cite{gilpin_explaining_2018,shmueli_explain_2010}.  Within the legal and philosophical traditions, there are existing ways of interpreting interpretability that have potentials for how students approach material technically.  An additional next step is to integrate existing curricular attempts to teach RDS, which overwhelmingly focus on humanistic approaches to the topic, and identify goals in common that allow us to begin to create a taxonomy of RDS pedagogy, and examine the effectiveness of ethical approaches in technical and humanistic courses.

%% file: main.bbl
%%% -*-BibTeX-*-
%%% Do NOT edit. File created by BibTeX with style
%%% ACM-Reference-Format-Journals [18-Jan-2012].

\begin{thebibliography}{113}

%%% ====================================================================
%%% NOTE TO THE USER: you can override these defaults by providing
%%% customized versions of any of these macros before the \bibliography
%%% command.  Each of them MUST provide its own final punctuation,
%%% except for \shownote{}, \showDOI{}, and \showURL{}.  The latter two
%%% do not use final punctuation, in order to avoid confusing it with
%%% the Web address.
%%%
%%% To suppress output of a particular field, define its macro to expand
%%% to an empty string, or better, \unskip, like this:
%%%
%%% \newcommand{\showDOI}[1]{\unskip}   % LaTeX syntax
%%%
%%% \def \showDOI #1{\unskip}           % plain TeX syntax
%%%
%%% ====================================================================

\ifx \showCODEN    \undefined \def \showCODEN     #1{\unskip}     \fi
\ifx \showDOI      \undefined \def \showDOI       #1{#1}\fi
\ifx \showISBNx    \undefined \def \showISBNx     #1{\unskip}     \fi
\ifx \showISBNxiii \undefined \def \showISBNxiii  #1{\unskip}     \fi
\ifx \showISSN     \undefined \def \showISSN      #1{\unskip}     \fi
\ifx \showLCCN     \undefined \def \showLCCN      #1{\unskip}     \fi
\ifx \shownote     \undefined \def \shownote      #1{#1}          \fi
\ifx \showarticletitle \undefined \def \showarticletitle #1{#1}   \fi
\ifx \showURL      \undefined \def \showURL       {\relax}        \fi
% The following commands are used for tagged output and should be
% invisible to TeX
\providecommand\bibfield[2]{#2}
\providecommand\bibinfo[2]{#2}
\providecommand\natexlab[1]{#1}
\providecommand\showeprint[2][]{arXiv:#2}

\bibitem[\protect\citeauthoryear{??}{op}{[n. d.]}]%
        {op}
 \bibinfo{year}{[n. d.]}\natexlab{}.
\newblock \bibinfo{title}{Open Provenance}.
\newblock \bibinfo{howpublished}{\url{https://openprovenance.org}}.
\newblock
\newblock
\shownote{[Online; accessed 14-August-2019].}


\bibitem[\protect\citeauthoryear{Aasheim, Williams, Rutner, and
  Gardiner}{Aasheim et~al\mbox{.}}{2015}]%
        {aasheim_data_2015}
\bibfield{author}{\bibinfo{person}{Cheryl~L. Aasheim}, \bibinfo{person}{Susan
  Williams}, \bibinfo{person}{Paige Rutner}, {and} \bibinfo{person}{Adrian
  Gardiner}.} \bibinfo{year}{2015}\natexlab{}.
\newblock \showarticletitle{Data {Analytics} vs. {Data} {Science}: {A} {Study}
  of {Similarities} and {Differences} in {Undergraduate} {Programs} {Based} on
  {Course} {Descriptions}}.
\newblock \bibinfo{journal}{\emph{Journal of Information Systems Education}}
  \bibinfo{volume}{26}, \bibinfo{number}{2} (\bibinfo{year}{2015}),
  \bibinfo{pages}{103--115}.
\newblock
\urldef\tempurl%
\url{http://jise.org/Volume26/n2/JISEv26n2p103.pdf}
\showURL{%
\tempurl}


\bibitem[\protect\citeauthoryear{Abedjan, Golab, and Naumann}{Abedjan
  et~al\mbox{.}}{2015}]%
        {DBLP:journals/vldb/AbedjanGN15}
\bibfield{author}{\bibinfo{person}{Ziawasch Abedjan}, \bibinfo{person}{Lukasz
  Golab}, {and} \bibinfo{person}{Felix Naumann}.}
  \bibinfo{year}{2015}\natexlab{}.
\newblock \showarticletitle{Profiling relational data: a survey}.
\newblock \bibinfo{journal}{\emph{{VLDB} J.}} \bibinfo{volume}{24},
  \bibinfo{number}{4} (\bibinfo{year}{2015}), \bibinfo{pages}{557--581}.
\newblock
\urldef\tempurl%
\url{https://doi.org/10.1007/s00778-015-0389-y}
\showDOI{\tempurl}


\bibitem[\protect\citeauthoryear{Ackermann}{Ackermann}{2001}]%
        {ackermann_piagets_2001}
\bibfield{author}{\bibinfo{person}{Edith Ackermann}.}
  \bibinfo{year}{2001}\natexlab{}.
\newblock \showarticletitle{Piaget’s {Constructivism}, {Papert}’s
  {Constructionism}: {What}’s the difference?}. In
  \bibinfo{booktitle}{\emph{Conference {Proceedings}}}, Vol.~\bibinfo{volume}{1
  and 2}. \bibinfo{address}{Geneva, Switzerland}, \bibinfo{pages}{85--94}.
\newblock


\bibitem[\protect\citeauthoryear{Administration}{Administration}{1994}]%
        {food_and_drug_administration_nutritional_1994}
\bibfield{author}{\bibinfo{person}{Food {and}~Drug Administration}.}
  \bibinfo{year}{1994}\natexlab{}.
\newblock \bibinfo{title}{Nutritional {Labeling} and {Education} {Act} ({NLEA})
  {Requirements} (8/94 - 2/95)}.
\newblock
\newblock
\urldef\tempurl%
\url{https://www.fda.gov/inspections-compliance-enforcement-and-criminal-investigations/inspection-guides/nutritional-labeling-and-education-act-nlea-requirements-894-295#GUIDE%20FOR%20REVIEW%20OF%20NUTRITION}
\showURL{%
\tempurl}


\bibitem[\protect\citeauthoryear{Aleven and Koedinger}{Aleven and
  Koedinger}{2002}]%
        {aleven_effective_2002}
\bibfield{author}{\bibinfo{person}{Vincent~A. Aleven} {and}
  \bibinfo{person}{Kenneth~R. Koedinger}.} \bibinfo{year}{2002}\natexlab{}.
\newblock \showarticletitle{An effective metacognitive strategy: {Learning} by
  doing and explaining with a computer-based {Cognitive} {Tutor}}.
\newblock \bibinfo{journal}{\emph{Cognitive Science}} \bibinfo{volume}{26},
  \bibinfo{number}{2} (\bibinfo{year}{2002}), \bibinfo{pages}{147--179}.
\newblock


\bibitem[\protect\citeauthoryear{Ananny and Crawford}{Ananny and
  Crawford}{2016}]%
        {ananny_seeing_2016}
\bibfield{author}{\bibinfo{person}{Mike Ananny} {and} \bibinfo{person}{Kate
  Crawford}.} \bibinfo{year}{2016}\natexlab{}.
\newblock \showarticletitle{Seeing without knowing: {Limitations} of the
  transparency ideal and its application to algorithmic accountability}.
\newblock \bibinfo{journal}{\emph{New Media \& Society}} \bibinfo{volume}{20},
  \bibinfo{number}{3} (\bibinfo{year}{2016}), \bibinfo{pages}{973--989}.
\newblock
\urldef\tempurl%
\url{https://doi.org/10.1177/1461444816676645}
\showDOI{\tempurl}


\bibitem[\protect\citeauthoryear{Anderson, Bowring, McCauley, Pothering, and
  Starr}{Anderson et~al\mbox{.}}{2014}]%
        {anderson_undergraduate_2014}
\bibfield{author}{\bibinfo{person}{Paul Anderson}, \bibinfo{person}{James
  Bowring}, \bibinfo{person}{Renée McCauley}, \bibinfo{person}{George
  Pothering}, {and} \bibinfo{person}{Christopher Starr}.}
  \bibinfo{year}{2014}\natexlab{}.
\newblock \showarticletitle{An undergraduate degree in data science: curriculum
  and a decade of implementation experience}. In
  \bibinfo{booktitle}{\emph{Proceedings of the 45th {ACM} technical symposium
  on {Computer} science education}}. \bibinfo{address}{Atlanta, Georgia},
  \bibinfo{pages}{145--150}.
\newblock
\urldef\tempurl%
\url{https://doi.org/10.1145/2538862.2538936}
\showDOI{\tempurl}


\bibitem[\protect\citeauthoryear{Angelo and Cross}{Angelo and Cross}{1993}]%
        {angelo_classroom_1993}
\bibfield{author}{\bibinfo{person}{Thomas~A. Angelo} {and}
  \bibinfo{person}{K.~Patricia Cross}.} \bibinfo{year}{1993}\natexlab{}.
\newblock \bibinfo{booktitle}{\emph{Classroom {Assessment} {Techniques}: {A}
  {Handbook} for {College} {Teachers}} (\bibinfo{edition}{2nd} ed.)}.
\newblock \bibinfo{publisher}{Jossey-Bass}, \bibinfo{address}{San Francisco,
  CA}.
\newblock


\bibitem[\protect\citeauthoryear{Angwin, Larson, Mattu, and Kirchner}{Angwin
  et~al\mbox{.}}{2016}]%
        {Angwin2016}
\bibfield{author}{\bibinfo{person}{Julia Angwin}, \bibinfo{person}{Jeff
  Larson}, \bibinfo{person}{Surya Mattu}, {and} \bibinfo{person}{Lauren
  Kirchner}.} \bibinfo{year}{2016}\natexlab{}.
\newblock \bibinfo{title}{Machine bias: There’s software used across the
  country to predict future criminals. And it’s biased against blacks.
  (ProPublica)}.
\newblock
\newblock
\urldef\tempurl%
\url{https://www.propublica.org/article/machine-bias-risk-assessments-in-criminal-sentencing}
\showURL{%
\tempurl}


\bibitem[\protect\citeauthoryear{Baker, Clarke-Midura, and Ocumpaugh}{Baker
  et~al\mbox{.}}{2016}]%
        {baker_toward_2016}
\bibfield{author}{\bibinfo{person}{Ryan~S. Baker}, \bibinfo{person}{Jody
  Clarke-Midura}, {and} \bibinfo{person}{Jaclyn Ocumpaugh}.}
  \bibinfo{year}{2016}\natexlab{}.
\newblock \showarticletitle{Toward general models of effective science inquiry
  in virtual performance assessments}.
\newblock \bibinfo{journal}{\emph{Journal of Computer Assisted Learning}}
  \bibinfo{volume}{32}, \bibinfo{number}{3} (\bibinfo{year}{2016}),
  \bibinfo{pages}{267--280}.
\newblock


\bibitem[\protect\citeauthoryear{Ben-Ari}{Ben-Ari}{2001}]%
        {ben-ari_constructivism_2001}
\bibfield{author}{\bibinfo{person}{Mordechai Ben-Ari}.}
  \bibinfo{year}{2001}\natexlab{}.
\newblock \showarticletitle{Constructivism in {Computer} {Science}
  {Education}}.
\newblock \bibinfo{journal}{\emph{Journal of Computers in Mathematics and
  Science Teaching}} \bibinfo{volume}{20}, \bibinfo{number}{1}
  (\bibinfo{year}{2001}), \bibinfo{pages}{45--73}.
\newblock


\bibitem[\protect\citeauthoryear{Biessmann, Salinas, Schelter, Schmidt, and
  Lange}{Biessmann et~al\mbox{.}}{2018}]%
        {Biessmann2018}
\bibfield{author}{\bibinfo{person}{Felix Biessmann}, \bibinfo{person}{David
  Salinas}, \bibinfo{person}{Sebastian Schelter}, \bibinfo{person}{Philipp
  Schmidt}, {and} \bibinfo{person}{Dustin Lange}.}
  \bibinfo{year}{2018}\natexlab{}.
\newblock \showarticletitle{Deep Learning for Missing Value Imputation in
  Tables with Non-Numerical Data}. In \bibinfo{booktitle}{\emph{Proceedings of
  the 27th ACM International Conference on Information and Knowledge
  Management}}. ACM, \bibinfo{pages}{2017--2025}.
\newblock


\bibitem[\protect\citeauthoryear{boyd}{boyd}{2010}]%
        {boyd_privacy_2010}
\bibfield{author}{\bibinfo{person}{danah boyd}.}
  \bibinfo{year}{2010}\natexlab{}.
\newblock \bibinfo{title}{Privacy and {Publicity} in the {Context} of {Big}
  {Data}}.
\newblock
\newblock
\urldef\tempurl%
\url{https://www.danah.org/papers/talks/2010/WWW2010.html}
\showURL{%
\tempurl}


\bibitem[\protect\citeauthoryear{Breiman}{Breiman}{2001}]%
        {breiman_statistical_2001}
\bibfield{author}{\bibinfo{person}{Leo Breiman}.}
  \bibinfo{year}{2001}\natexlab{}.
\newblock \showarticletitle{Statistical {Modeling}: {The} {Two} {Cultures}}.
\newblock \bibinfo{journal}{\emph{Statist. Sci.}} \bibinfo{volume}{16},
  \bibinfo{number}{3} (\bibinfo{year}{2001}), \bibinfo{pages}{199--231}.
\newblock
\urldef\tempurl%
\url{https://doi.org/10.1214/ss/1009213726}
\showDOI{\tempurl}


\bibitem[\protect\citeauthoryear{Broussard}{Broussard}{2018}]%
        {broussard_artificial_2018}
\bibfield{author}{\bibinfo{person}{Meredith Broussard}.}
  \bibinfo{year}{2018}\natexlab{}.
\newblock \bibinfo{booktitle}{\emph{Artificial {Unintelligence}: How
  {Computers} {Misunderstand} the {World}}}.
\newblock \bibinfo{publisher}{MIT Press}, \bibinfo{address}{Cambridge,
  Massachusetts}.
\newblock
\showISBNx{9780262038003}


\bibitem[\protect\citeauthoryear{Byrd-Bredbenner, Alfieri, Wong, and
  Cottee}{Byrd-Bredbenner et~al\mbox{.}}{2009}]%
        {byrd-bredbenner_inherent_2009}
\bibfield{author}{\bibinfo{person}{Carol Byrd-Bredbenner},
  \bibinfo{person}{Lisa Alfieri}, \bibinfo{person}{Angela Wong}, {and}
  \bibinfo{person}{Peta Cottee}.} \bibinfo{year}{2009}\natexlab{}.
\newblock \showarticletitle{The inherent educational qualities of nutrition
  labels}.
\newblock \bibinfo{journal}{\emph{Family \& Consumer Sciences Research
  Journal}} \bibinfo{volume}{29}, \bibinfo{number}{26} (\bibinfo{year}{2009}).
\newblock
\urldef\tempurl%
\url{https://doi.org/10.1177/1077727X01293004}
\showDOI{\tempurl}


\bibitem[\protect\citeauthoryear{Carmichael and Marron}{Carmichael and
  Marron}{2018}]%
        {carmichael_data_2018}
\bibfield{author}{\bibinfo{person}{Iain Carmichael} {and} \bibinfo{person}{J.S.
  Marron}.} \bibinfo{year}{2018}\natexlab{}.
\newblock \showarticletitle{Data science vs. statistics: two cultures?}
\newblock \bibinfo{journal}{\emph{Japanese Journal of Statistics and Data
  Science}} \bibinfo{volume}{1}, \bibinfo{number}{1} (\bibinfo{year}{2018}),
  \bibinfo{pages}{117--138}.
\newblock
\urldef\tempurl%
\url{https://doi.org/10.1007/s42081-018-0009-3}
\showDOI{\tempurl}


\bibitem[\protect\citeauthoryear{Chamberlin and Powers}{Chamberlin and
  Powers}{2010}]%
        {chamberlin_promise_2010}
\bibfield{author}{\bibinfo{person}{Michelle Chamberlin} {and}
  \bibinfo{person}{Robert Powers}.} \bibinfo{year}{2010}\natexlab{}.
\newblock \showarticletitle{The promise of differentiated instruction for
  enhancing the mathematical understandings of college students}.
\newblock \bibinfo{journal}{\emph{Teaching Mathematics and its Applications}}
  \bibinfo{volume}{29}, \bibinfo{number}{3} (\bibinfo{year}{2010}),
  \bibinfo{pages}{113--139}.
\newblock
\urldef\tempurl%
\url{https://doi.org/10.1093/teamat/hrq006}
\showDOI{\tempurl}


\bibitem[\protect\citeauthoryear{Chouldechova}{Chouldechova}{2017}]%
        {DBLP:journals/corr/Chouldechova17}
\bibfield{author}{\bibinfo{person}{Alexandra Chouldechova}.}
  \bibinfo{year}{2017}\natexlab{}.
\newblock \showarticletitle{Fair prediction with disparate impact: {A} study of
  bias in recidivism prediction instruments}.
\newblock \bibinfo{journal}{\emph{CoRR}}  \bibinfo{volume}{abs/1703.00056}
  (\bibinfo{year}{2017}).
\newblock
\showeprint[arxiv]{1703.00056}
\urldef\tempurl%
\url{http://arxiv.org/abs/1703.00056}
\showURL{%
\tempurl}


\bibitem[\protect\citeauthoryear{Chu and Ilyas}{Chu and Ilyas}{2016}]%
        {DBLP:journals/pvldb/ChuI16}
\bibfield{author}{\bibinfo{person}{Xu Chu} {and} \bibinfo{person}{Ihab~F.
  Ilyas}.} \bibinfo{year}{2016}\natexlab{}.
\newblock \showarticletitle{Qualitative Data Cleaning}.
\newblock \bibinfo{journal}{\emph{{PVLDB}}} \bibinfo{volume}{9},
  \bibinfo{number}{13} (\bibinfo{year}{2016}), \bibinfo{pages}{1605--1608}.
\newblock
\urldef\tempurl%
\url{https://doi.org/10.14778/3007263.3007320}
\showDOI{\tempurl}


\bibitem[\protect\citeauthoryear{Committee}{Committee}{2016}]%
        {gaise_college_report_asa_revision_committee_guidelines_2016}
\bibfield{author}{\bibinfo{person}{GAISE College Report ASA~Revision
  Committee}.} \bibinfo{year}{2016}\natexlab{}.
\newblock \bibinfo{title}{Guidelines for assessment and instruction in
  statistics education ({GAISE}): {College} report 2016}.
\newblock
\newblock
\urldef\tempurl%
\url{http://www.amstat.org/education/gaise}
\showURL{%
\tempurl}


\bibitem[\protect\citeauthoryear{Council}{Council}{2000}]%
        {national_research_council_how_2000}
\bibfield{author}{\bibinfo{person}{National~Research Council}.}
  \bibinfo{year}{2000}\natexlab{}.
\newblock \bibinfo{booktitle}{\emph{How {People} {Learn}: {Brain}, {Mind},
  {Experience}, and {School}} (\bibinfo{edition}{expanded edition} ed.)}.
\newblock \bibinfo{publisher}{The National Academies Press},
  \bibinfo{address}{Washington, DC}.
\newblock
\showISBNx{978-0-309-07036-2}


\bibitem[\protect\citeauthoryear{Datta, Datta, Makagon, Mulligan, and
  Tschantz}{Datta et~al\mbox{.}}{2018}]%
        {DBLP:conf/fat/DattaDMMT18}
\bibfield{author}{\bibinfo{person}{Amit Datta}, \bibinfo{person}{Anupam Datta},
  \bibinfo{person}{Jael Makagon}, \bibinfo{person}{Deirdre~K. Mulligan}, {and}
  \bibinfo{person}{Michael~Carl Tschantz}.} \bibinfo{year}{2018}\natexlab{}.
\newblock \showarticletitle{Discrimination in Online Personalization: {A}
  Multidisciplinary Inquiry}. In \bibinfo{booktitle}{\emph{Conference on
  Fairness, Accountability and Transparency, {FAT} 2018, 23-24 February 2018,
  New York, NY, {USA}}}. \bibinfo{pages}{20--34}.
\newblock
\urldef\tempurl%
\url{http://proceedings.mlr.press/v81/datta18a.html}
\showURL{%
\tempurl}


\bibitem[\protect\citeauthoryear{Datta, Sen, and Zick}{Datta
  et~al\mbox{.}}{2016}]%
        {datta_algorithmic_2016}
\bibfield{author}{\bibinfo{person}{Anupam Datta}, \bibinfo{person}{Shayak Sen},
  {and} \bibinfo{person}{Yair Zick}.} \bibinfo{year}{2016}\natexlab{}.
\newblock \showarticletitle{Algorithmic transparency via quantitative input
  influence: {Theory} and experiments with learning systems}. In
  \bibinfo{booktitle}{\emph{Proceedings of {IEEE} {Symposium} on {Security} and
  {Privacy}}}. \bibinfo{publisher}{IEEE}, \bibinfo{pages}{598--617}.
\newblock


\bibitem[\protect\citeauthoryear{Datta, Tschantz, and Datta}{Datta
  et~al\mbox{.}}{2015}]%
        {DBLP:journals/popets/DattaTD15}
\bibfield{author}{\bibinfo{person}{Amit Datta}, \bibinfo{person}{Michael~Carl
  Tschantz}, {and} \bibinfo{person}{Anupam Datta}.}
  \bibinfo{year}{2015}\natexlab{}.
\newblock \showarticletitle{Automated Experiments on Ad Privacy Settings}.
\newblock \bibinfo{journal}{\emph{PoPETs}} \bibinfo{volume}{2015},
  \bibinfo{number}{1} (\bibinfo{year}{2015}), \bibinfo{pages}{92--112}.
\newblock
\urldef\tempurl%
\url{https://doi.org/10.1515/popets-2015-0007}
\showDOI{\tempurl}


\bibitem[\protect\citeauthoryear{De~Veaux, Agarwal, Averett, Baumer, Bray,
  Bressoud, Bryant, Cheng, Francis, Gould, Kim, Kretchmar, Lu, Moskol, Nolan,
  Pelayo, Raleigh, Sethi, Sondjaja, Tiruviluamala, Uhlig, Washington, Wesley,
  White, and Ye}{De~Veaux et~al\mbox{.}}{2017}]%
        {de_veaux_curriculum_2017}
\bibfield{author}{\bibinfo{person}{Richard De~Veaux}, \bibinfo{person}{Mahesh
  Agarwal}, \bibinfo{person}{Maia Averett}, \bibinfo{person}{Benjamin~S.
  Baumer}, \bibinfo{person}{Andrew Bray}, \bibinfo{person}{Thomas~C. Bressoud},
  \bibinfo{person}{Lance Bryant}, \bibinfo{person}{Lei~Z. Cheng},
  \bibinfo{person}{Amanda Francis}, \bibinfo{person}{Robert Gould},
  \bibinfo{person}{Albert~Y. Kim}, \bibinfo{person}{Matt Kretchmar},
  \bibinfo{person}{Qin Lu}, \bibinfo{person}{Ann Moskol},
  \bibinfo{person}{Deborah Nolan}, \bibinfo{person}{Roberto Pelayo},
  \bibinfo{person}{Sean Raleigh}, \bibinfo{person}{Ricky~J. Sethi},
  \bibinfo{person}{Mutiara Sondjaja}, \bibinfo{person}{Neelesh Tiruviluamala},
  \bibinfo{person}{Paul~X. Uhlig}, \bibinfo{person}{Talitha~M. Washington},
  \bibinfo{person}{Curtis~L. Wesley}, \bibinfo{person}{David White}, {and}
  \bibinfo{person}{Ping Ye}.} \bibinfo{year}{2017}\natexlab{}.
\newblock \showarticletitle{Curriculum {Guidelines} for {Undergraduate}
  {Programs} in {Data} {Science}}.
\newblock \bibinfo{journal}{\emph{Annual Review of Statistics and Its
  Application}}  \bibinfo{volume}{4} (\bibinfo{year}{2017}),
  \bibinfo{pages}{2.1--2.16}.
\newblock
\urldef\tempurl%
\url{https://doi.org/10.1146/annurev-statistics-060116-053930}
\showDOI{\tempurl}


\bibitem[\protect\citeauthoryear{Diakopoulos}{Diakopoulos}{2016}]%
        {diakopoulos_accountability_2016}
\bibfield{author}{\bibinfo{person}{Nicholas Diakopoulos}.}
  \bibinfo{year}{2016}\natexlab{}.
\newblock \showarticletitle{Accountability in algorithmic decision making}.
\newblock \bibinfo{journal}{\emph{Commun. ACM}} \bibinfo{volume}{59},
  \bibinfo{number}{2} (\bibinfo{date}{Feb.} \bibinfo{year}{2016}),
  \bibinfo{pages}{56--62}.
\newblock
\urldef\tempurl%
\url{https://doi.org/10.1145/2844110}
\showDOI{\tempurl}


\bibitem[\protect\citeauthoryear{Dinur and Nissim}{Dinur and Nissim}{2003}]%
        {DBLP:conf/pods/DinurN03}
\bibfield{author}{\bibinfo{person}{Irit Dinur} {and} \bibinfo{person}{Kobbi
  Nissim}.} \bibinfo{year}{2003}\natexlab{}.
\newblock \showarticletitle{Revealing information while preserving privacy}. In
  \bibinfo{booktitle}{\emph{Proceedings of the Twenty-Second {ACM}
  {SIGACT-SIGMOD-SIGART} Symposium on Principles of Database Systems, June
  9-12, 2003, San Diego, CA, {USA}}}. \bibinfo{pages}{202--210}.
\newblock
\urldef\tempurl%
\url{https://doi.org/10.1145/773153.773173}
\showDOI{\tempurl}


\bibitem[\protect\citeauthoryear{diSessa, Forman, and Pufall}{diSessa
  et~al\mbox{.}}{1988}]%
        {disessa_knowledge_1988}
\bibfield{author}{\bibinfo{person}{Andrea~A. diSessa}, \bibinfo{person}{George
  Forman}, {and} \bibinfo{person}{Peter~P. Pufall}.}
  \bibinfo{year}{1988}\natexlab{}.
\newblock \showarticletitle{Knowledge in pieces}.
\newblock In \bibinfo{booktitle}{\emph{Constructivism in the {Computer}
  {Age}}}. \bibinfo{publisher}{Lawrence Erlbaum}, \bibinfo{address}{Hillsdale,
  NJ}, \bibinfo{pages}{49--70}.
\newblock


\bibitem[\protect\citeauthoryear{Doshi-Velez and Kim}{Doshi-Velez and
  Kim}{2017}]%
        {doshi-velez_towards_2017}
\bibfield{author}{\bibinfo{person}{Finale Doshi-Velez} {and}
  \bibinfo{person}{Been Kim}.} \bibinfo{year}{2017}\natexlab{}.
\newblock \showarticletitle{Towards a rigorous science of interpretable machine
  learning}.
\newblock  (\bibinfo{year}{2017}).
\newblock
\urldef\tempurl%
\url{https://arxiv.org/pdf/1702.08608.pdf}
\showURL{%
\tempurl}


\bibitem[\protect\citeauthoryear{Dweck}{Dweck}{1986}]%
        {dweck_motivational_1986}
\bibfield{author}{\bibinfo{person}{Carol~S. Dweck}.}
  \bibinfo{year}{1986}\natexlab{}.
\newblock \showarticletitle{Motivational processes affect learning}.
\newblock \bibinfo{journal}{\emph{American Psychologist}}  \bibinfo{volume}{41}
  (\bibinfo{year}{1986}), \bibinfo{pages}{1040--1048}.
\newblock


\bibitem[\protect\citeauthoryear{Dwork}{Dwork}{2011}]%
        {DBLP:journals/cacm/Dwork11}
\bibfield{author}{\bibinfo{person}{Cynthia Dwork}.}
  \bibinfo{year}{2011}\natexlab{}.
\newblock \showarticletitle{A firm foundation for private data analysis}.
\newblock \bibinfo{journal}{\emph{Commun. {ACM}}} \bibinfo{volume}{54},
  \bibinfo{number}{1} (\bibinfo{year}{2011}), \bibinfo{pages}{86--95}.
\newblock
\urldef\tempurl%
\url{https://doi.org/10.1145/1866739.1866758}
\showDOI{\tempurl}


\bibitem[\protect\citeauthoryear{Dwork, Hardt, Pitassi, Reingold, and
  Zemel}{Dwork et~al\mbox{.}}{2012}]%
        {DBLP:conf/innovations/DworkHPRZ12}
\bibfield{author}{\bibinfo{person}{Cynthia Dwork}, \bibinfo{person}{Moritz
  Hardt}, \bibinfo{person}{Toniann Pitassi}, \bibinfo{person}{Omer Reingold},
  {and} \bibinfo{person}{Richard~S. Zemel}.} \bibinfo{year}{2012}\natexlab{}.
\newblock \showarticletitle{Fairness through awareness}. In
  \bibinfo{booktitle}{\emph{Innovations in Theoretical Computer Science 2012,
  Cambridge, MA, USA, January 8-10, 2012}}. \bibinfo{pages}{214--226}.
\newblock
\urldef\tempurl%
\url{https://doi.org/10.1145/2090236.2090255}
\showDOI{\tempurl}


\bibitem[\protect\citeauthoryear{Farahi and Stroud}{Farahi and Stroud}{2018}]%
        {farahi_michigan_2018}
\bibfield{author}{\bibinfo{person}{Arya Farahi} {and} \bibinfo{person}{Jonathan
  Stroud}.} \bibinfo{year}{2018}\natexlab{}.
\newblock \showarticletitle{The {Michigan} {Data} {Science} {Team}: {A} {Data}
  {Science} {Education} {Program} with {Significant} {Social} {Impact}}. In
  \bibinfo{booktitle}{\emph{Proceedings of 2018 {DSW}}}.
  \bibinfo{address}{Lausanne, Switzerland}.
\newblock


\bibitem[\protect\citeauthoryear{for Computing~Machinery}{for
  Computing~Machinery}{2018}]%
        {association_for_computing_machinery_acm_2018}
\bibfield{author}{\bibinfo{person}{Association for Computing~Machinery}.}
  \bibinfo{year}{2018}\natexlab{}.
\newblock \bibinfo{title}{{ACM} code of ethics and professional conduct}.
\newblock
\newblock
\urldef\tempurl%
\url{https://www.acm.org/code-of-ethics}
\showURL{%
\tempurl}


\bibitem[\protect\citeauthoryear{Friedler, Scheidegger, and
  Venkatasubramanian}{Friedler et~al\mbox{.}}{2016}]%
        {DBLP:journals/corr/FriedlerSV16}
\bibfield{author}{\bibinfo{person}{Sorelle~A. Friedler},
  \bibinfo{person}{Carlos Scheidegger}, {and} \bibinfo{person}{Suresh
  Venkatasubramanian}.} \bibinfo{year}{2016}\natexlab{}.
\newblock \showarticletitle{On the (im)possibility of fairness}.
\newblock \bibinfo{journal}{\emph{CoRR}}  \bibinfo{volume}{abs/1609.07236}
  (\bibinfo{year}{2016}).
\newblock
\showeprint[arxiv]{1609.07236}
\urldef\tempurl%
\url{http://arxiv.org/abs/1609.07236}
\showURL{%
\tempurl}


\bibitem[\protect\citeauthoryear{Friedman and Winograd}{Friedman and
  Winograd}{1990}]%
        {friedman_computing_1990}
\bibfield{author}{\bibinfo{person}{Batya Friedman} {and} \bibinfo{person}{Terry
  Winograd}.} \bibinfo{year}{1990}\natexlab{}.
\newblock \bibinfo{booktitle}{\emph{Computing and social responsibility: a
  collection of course syllabi}}.
\newblock \bibinfo{publisher}{Computer Professionals for Social
  Responsibility}, \bibinfo{address}{Palo Alto, CA}.
\newblock
\urldef\tempurl%
\url{https://dl.acm.org/citation.cfm?id=152167}
\showURL{%
\tempurl}


\bibitem[\protect\citeauthoryear{Garfield and Ben-Zvi}{Garfield and
  Ben-Zvi}{2007}]%
        {garfield_how_2007}
\bibfield{author}{\bibinfo{person}{Joan Garfield} {and} \bibinfo{person}{Dani
  Ben-Zvi}.} \bibinfo{year}{2007}\natexlab{}.
\newblock \showarticletitle{How {Students} {Learn} {Statistics} {Revisited}:
  {A} {Current} {Review} of {Research} on {Teaching} and {Learning}
  {Statistics}}.
\newblock \bibinfo{journal}{\emph{International Statistical Review}}
  \bibinfo{volume}{75}, \bibinfo{number}{3} (\bibinfo{year}{2007}).
\newblock
\urldef\tempurl%
\url{https://doi.org/10.1111/j.1751-5823.2007.00029.x}
\showDOI{\tempurl}


\bibitem[\protect\citeauthoryear{Gebru, Morgenstern, Vecchione, Vaughan,
  Wallach, Daume{\'e}~III, and Crawford}{Gebru et~al\mbox{.}}{2018}]%
        {Gebru2018}
\bibfield{author}{\bibinfo{person}{Timnit Gebru}, \bibinfo{person}{Jamie
  Morgenstern}, \bibinfo{person}{Briana Vecchione},
  \bibinfo{person}{Jennifer~Wortman Vaughan}, \bibinfo{person}{Hanna Wallach},
  \bibinfo{person}{Hal Daume{\'e}~III}, {and} \bibinfo{person}{Kate Crawford}.}
  \bibinfo{year}{2018}\natexlab{}.
\newblock \showarticletitle{Datasheets for Datasets}.
\newblock \bibinfo{journal}{\emph{arXiv preprint arXiv:1803.09010}}
  (\bibinfo{year}{2018}).
\newblock


\bibitem[\protect\citeauthoryear{George}{George}{2005}]%
        {george_rationale_2005}
\bibfield{author}{\bibinfo{person}{Paul~S. George}.}
  \bibinfo{year}{2005}\natexlab{}.
\newblock \showarticletitle{A rationale for differentiating instruction in the
  regular classroom}.
\newblock \bibinfo{journal}{\emph{Theory Into Practice}} \bibinfo{volume}{44},
  \bibinfo{number}{3} (\bibinfo{year}{2005}), \bibinfo{pages}{185--193}.
\newblock
\urldef\tempurl%
\url{https://doi.org/10.1207/s15430421tip4403_2}
\showDOI{\tempurl}


\bibitem[\protect\citeauthoryear{Gillborn, Warmington, and Demack}{Gillborn
  et~al\mbox{.}}{2017}]%
        {gillborn_quantcrit:_2017}
\bibfield{author}{\bibinfo{person}{David Gillborn}, \bibinfo{person}{Paul
  Warmington}, {and} \bibinfo{person}{Sean Demack}.}
  \bibinfo{year}{2017}\natexlab{}.
\newblock \showarticletitle{{QuantCrit}: education, policy, ‘{Big} {Data}’
  and principles for a critical race theory of statistics}.
\newblock \bibinfo{journal}{\emph{Race, Ethnicity, and Education}}
  \bibinfo{volume}{2} (\bibinfo{year}{2017}), \bibinfo{pages}{158--179}.
\newblock
\urldef\tempurl%
\url{https://doi.org/10.1080/13613324.2017.1377417}
\showDOI{\tempurl}


\bibitem[\protect\citeauthoryear{Gilpin, Bau, Yuan, Bajwa, Specter, and
  Kagal}{Gilpin et~al\mbox{.}}{2018}]%
        {gilpin_explaining_2018}
\bibfield{author}{\bibinfo{person}{Leilani~H. Gilpin}, \bibinfo{person}{David
  Bau}, \bibinfo{person}{Ben~Z. Yuan}, \bibinfo{person}{Ayesha Bajwa},
  \bibinfo{person}{Michael Specter}, {and} \bibinfo{person}{Lalana Kagal}.}
  \bibinfo{year}{2018}\natexlab{}.
\newblock \showarticletitle{Explaining {Explanations}: {An} {Overview} of
  {Interpretability} of {Machine} {Learning}}. In
  \bibinfo{booktitle}{\emph{{DSAA} 2018}}.
\newblock
\urldef\tempurl%
\url{https://arxiv.org/pdf/1806.00069.pdf}
\showURL{%
\tempurl}


\bibitem[\protect\citeauthoryear{Gitelman and Jackson}{Gitelman and
  Jackson}{2013}]%
        {gitelman_introduction_2013}
\bibfield{author}{\bibinfo{person}{Lisa Gitelman} {and}
  \bibinfo{person}{Virginia Jackson}.} \bibinfo{year}{2013}\natexlab{}.
\newblock \showarticletitle{Introduction}.
\newblock In \bibinfo{booktitle}{\emph{"{Raw} {Data}" {Is} an {Oxymoron}}},
  \bibfield{editor}{\bibinfo{person}{Lisa Gitelman}} (Ed.).
  \bibinfo{publisher}{MIT Press}, \bibinfo{address}{Cambridge, MA},
  \bibinfo{pages}{1--14}.
\newblock
\showISBNx{978-0262518284}


\bibitem[\protect\citeauthoryear{Gleicher}{Gleicher}{2016}]%
        {gleicher_framework_2016}
\bibfield{author}{\bibinfo{person}{Michael Gleicher}.}
  \bibinfo{year}{2016}\natexlab{}.
\newblock \showarticletitle{A {Framework} for {Considering} {Comprehensibility}
  in {Modeling}}.
\newblock \bibinfo{journal}{\emph{Big Data}} \bibinfo{volume}{4},
  \bibinfo{number}{2} (\bibinfo{year}{2016}), \bibinfo{pages}{75--88}.
\newblock
\urldef\tempurl%
\url{https://doi.org/10.1089/big.2016.0007}
\showDOI{\tempurl}


\bibitem[\protect\citeauthoryear{Goodfellow, Shlens, and Szegedy}{Goodfellow
  et~al\mbox{.}}{2015}]%
        {goodfellow_explaining_2015}
\bibfield{author}{\bibinfo{person}{Ian~J. Goodfellow},
  \bibinfo{person}{Jonathan Shlens}, {and} \bibinfo{person}{Christian
  Szegedy}.} \bibinfo{year}{2015}\natexlab{}.
\newblock \showarticletitle{Explaining and harnessing adversarial examples}. In
  \bibinfo{booktitle}{\emph{Proceedings from the {International} {Conference}
  on {Learning} {Representations} 2015}}. \bibinfo{address}{San Diego, CA},
  \bibinfo{pages}{1--11}.
\newblock
\urldef\tempurl%
\url{https://arxiv.org/pdf/1412.6572.pdf}
\showURL{%
\tempurl}


\bibitem[\protect\citeauthoryear{Goodman and Flaxman}{Goodman and
  Flaxman}{2017}]%
        {goodman_european_2017}
\bibfield{author}{\bibinfo{person}{Bryce Goodman} {and} \bibinfo{person}{Seth
  Flaxman}.} \bibinfo{year}{2017}\natexlab{}.
\newblock \showarticletitle{European {Union} {Regulations} on {Algorithmic}
  {Decision}-{Making} and a “{Right} to {Explanation}”}.
\newblock \bibinfo{journal}{\emph{AI Magazine}} \bibinfo{volume}{38},
  \bibinfo{number}{3} (\bibinfo{year}{2017}), \bibinfo{pages}{50--57}.
\newblock
\urldef\tempurl%
\url{https://doi.org/10.1609/aimag.v38i3.2741}
\showURL{%
\tempurl}


\bibitem[\protect\citeauthoryear{Guidotti, Monreale, Ruggieri, Turini,
  Pedreschi, and Giannotti}{Guidotti et~al\mbox{.}}{2018}]%
        {guidotti_survey_2018}
\bibfield{author}{\bibinfo{person}{Riccardo Guidotti}, \bibinfo{person}{Anna
  Monreale}, \bibinfo{person}{Salvatore Ruggieri}, \bibinfo{person}{Franco
  Turini}, \bibinfo{person}{Dino Pedreschi}, {and} \bibinfo{person}{Fosca
  Giannotti}.} \bibinfo{year}{2018}\natexlab{}.
\newblock \showarticletitle{A {Survey} of {Methods} for {Explaining} {Black}
  {Box} {Models}}.
\newblock \bibinfo{journal}{\emph{ACM Computing Surveys (CSUR) Surveys}}
  \bibinfo{volume}{51}, \bibinfo{number}{5} (\bibinfo{year}{2018}).
\newblock
\urldef\tempurl%
\url{https://doi.org/10.1145/3236009}
\showDOI{\tempurl}


\bibitem[\protect\citeauthoryear{Gunaratne and Nov}{Gunaratne and Nov}{2017}]%
        {gunaratne_using_2017}
\bibfield{author}{\bibinfo{person}{Junius Gunaratne} {and}
  \bibinfo{person}{Oded Nov}.} \bibinfo{year}{2017}\natexlab{}.
\newblock \showarticletitle{Using {Interactive} "{Nutrition} {Labels}" for
  {Financial} {Products} to {Assist} {Decision} {Making} Under {Uncertainty}}.
\newblock \bibinfo{journal}{\emph{Journal of the Association for Information
  Science and Technology}} \bibinfo{volume}{68}, \bibinfo{number}{8}
  (\bibinfo{year}{2017}), \bibinfo{pages}{1836--1849}.
\newblock
\urldef\tempurl%
\url{https://doi.org/10.1002/asi.23844}
\showDOI{\tempurl}


\bibitem[\protect\citeauthoryear{Hilliam and Calvert}{Hilliam and
  Calvert}{2019}]%
        {hilliam_interactive_2019}
\bibfield{author}{\bibinfo{person}{Rachel Hilliam} {and} \bibinfo{person}{Carol
  Calvert}.} \bibinfo{year}{2019}\natexlab{}.
\newblock \showarticletitle{Interactive statistics for a diverse student
  population}.
\newblock \bibinfo{journal}{\emph{Open Learning: The Journal of Open, Distance
  and e-Learning}} \bibinfo{volume}{34}, \bibinfo{number}{2}
  (\bibinfo{year}{2019}).
\newblock


\bibitem[\protect\citeauthoryear{Holland, Hosny, Newman, Joseph, and
  Chmielinski}{Holland et~al\mbox{.}}{2018}]%
        {DBLP:journals/corr/abs-1805-03677}
\bibfield{author}{\bibinfo{person}{Sarah Holland}, \bibinfo{person}{Ahmed
  Hosny}, \bibinfo{person}{Sarah Newman}, \bibinfo{person}{Joshua Joseph},
  {and} \bibinfo{person}{Kasia Chmielinski}.} \bibinfo{year}{2018}\natexlab{}.
\newblock \showarticletitle{The Dataset Nutrition Label: {A} Framework To Drive
  Higher Data Quality Standards}.
\newblock \bibinfo{journal}{\emph{CoRR}}  \bibinfo{volume}{abs/1805.03677}
  (\bibinfo{year}{2018}).
\newblock
\showeprint[arxiv]{1805.03677}
\urldef\tempurl%
\url{http://arxiv.org/abs/1805.03677}
\showURL{%
\tempurl}


\bibitem[\protect\citeauthoryear{Homelessness}{Homelessness}{[n. d.]}]%
        {VISPDAT}
\bibfield{author}{\bibinfo{person}{Partners~Ending Homelessness}.}
  \bibinfo{year}{[n. d.]}\natexlab{}.
\newblock \bibinfo{title}{{Vulnerability Index - Service Prioritization
  Decision Assistance Tool (VI-SPDAT)}}.
\newblock
  \bibinfo{howpublished}{\url{http://pehgc.org/wp-content/uploads/2016/09/VI-SPDAT-v2.01-Single-US-Fillable.pdf}}.
\newblock
\newblock
\shownote{[Online; accessed on 14-September-2017].}


\bibitem[\protect\citeauthoryear{Huff and Martin}{Huff and Martin}{1995}]%
        {huff_computing_1995}
\bibfield{author}{\bibinfo{person}{Chuck Huff} {and} \bibinfo{person}{C.~Dianne
  Martin}.} \bibinfo{year}{1995}\natexlab{}.
\newblock \showarticletitle{Computing consequences: a framework for teaching
  ethical computing}.
\newblock \bibinfo{journal}{\emph{Commun. ACM}} \bibinfo{volume}{38},
  \bibinfo{number}{12} (\bibinfo{date}{Dec.} \bibinfo{year}{1995}),
  \bibinfo{pages}{75--84}.
\newblock
\urldef\tempurl%
\url{https://doi.org/10.1145/219663.219687}
\showDOI{\tempurl}


\bibitem[\protect\citeauthoryear{Hundhausen and Douglas}{Hundhausen and
  Douglas}{2000}]%
        {hundhausen_using_2000}
\bibfield{author}{\bibinfo{person}{Christopher~D. Hundhausen} {and}
  \bibinfo{person}{Sarah~A. Douglas}.} \bibinfo{year}{2000}\natexlab{}.
\newblock \showarticletitle{Using visualizations to learn algorithms: should
  students construct their own, or view an expert’s?}. In
  \bibinfo{booktitle}{\emph{Proceedings of 2000 {IEEE} {Symposiumon} on
  {Visual} {Languages}}}. \bibinfo{publisher}{IEEE Computer Society Press},
  \bibinfo{address}{Los Alamitos, CA}, \bibinfo{pages}{21--28}.
\newblock


\bibitem[\protect\citeauthoryear{Hundhausen, Douglas, and Stasko}{Hundhausen
  et~al\mbox{.}}{2002}]%
        {hundhausen_meta-study_2002}
\bibfield{author}{\bibinfo{person}{Christopher~D. Hundhausen},
  \bibinfo{person}{Sarah~A. Douglas}, {and} \bibinfo{person}{John~T. Stasko}.}
  \bibinfo{year}{2002}\natexlab{}.
\newblock \showarticletitle{A {Meta}-{Study} of {Algorithm} {Visualization}
  {E}ffectiveness}.
\newblock \bibinfo{journal}{\emph{Journal of Visual Languages and Computing}}
  \bibinfo{volume}{13} (\bibinfo{year}{2002}), \bibinfo{pages}{259--290}.
\newblock
\urldef\tempurl%
\url{https://doi.org/0.1006/S1045-926X(02)00028-9}
\showDOI{\tempurl}


\bibitem[\protect\citeauthoryear{Kelley, Cesca, Bresee, and Cranor}{Kelley
  et~al\mbox{.}}{2010}]%
        {kelley_standardizing_2010}
\bibfield{author}{\bibinfo{person}{Patrick~G. Kelley}, \bibinfo{person}{Lucian
  Cesca}, \bibinfo{person}{Joanna Bresee}, {and} \bibinfo{person}{Lorrie~F.
  Cranor}.} \bibinfo{year}{2010}\natexlab{}.
\newblock \showarticletitle{Standardizing privacy notices: {An} online study of
  the nutrition label approach}. In \bibinfo{booktitle}{\emph{Proceedings of
  the {SIGCHI} {Conference} on {Human} {Factors} in {Computing} {Systems}}}.
  \bibinfo{publisher}{ACM}, \bibinfo{address}{Atlanta, Georgia},
  \bibinfo{pages}{1573--1582}.
\newblock
\urldef\tempurl%
\url{https://doi.org/10.1145/1753326.1753561}
\showDOI{\tempurl}


\bibitem[\protect\citeauthoryear{Kim, Patel, Rostamizadeh, and Shah}{Kim
  et~al\mbox{.}}{2015}]%
        {kim_scalable_2015}
\bibfield{author}{\bibinfo{person}{Been Kim}, \bibinfo{person}{Kayur Patel},
  \bibinfo{person}{Afshin Rostamizadeh}, {and} \bibinfo{person}{Julie Shah}.}
  \bibinfo{year}{2015}\natexlab{}.
\newblock \showarticletitle{Scalable and interpretable data representation for
  high-dimensional, complex data}. In \bibinfo{booktitle}{\emph{Proceedings of
  {AAAI} 2015}}. \bibinfo{publisher}{AAAI}.
\newblock


\bibitem[\protect\citeauthoryear{Kirkpatrick}{Kirkpatrick}{2017}]%
        {Kirkpatrick:2017:AD:3042068.3022181}
\bibfield{author}{\bibinfo{person}{Keith Kirkpatrick}.}
  \bibinfo{year}{2017}\natexlab{}.
\newblock \showarticletitle{It's Not the Algorithm, It's the Data}.
\newblock \bibinfo{journal}{\emph{Commun. ACM}} \bibinfo{volume}{60},
  \bibinfo{number}{2} (\bibinfo{date}{Jan.} \bibinfo{year}{2017}),
  \bibinfo{pages}{21--23}.
\newblock
\showISSN{0001-0782}
\urldef\tempurl%
\url{https://doi.org/10.1145/3022181}
\showDOI{\tempurl}


\bibitem[\protect\citeauthoryear{Kleinberg, Mullainathan, and
  Raghavan}{Kleinberg et~al\mbox{.}}{2017}]%
        {DBLP:conf/innovations/KleinbergMR17}
\bibfield{author}{\bibinfo{person}{Jon~M. Kleinberg}, \bibinfo{person}{Sendhil
  Mullainathan}, {and} \bibinfo{person}{Manish Raghavan}.}
  \bibinfo{year}{2017}\natexlab{}.
\newblock \showarticletitle{Inherent Trade-Offs in the Fair Determination of
  Risk Scores}. In \bibinfo{booktitle}{\emph{8th Innovations in Theoretical
  Computer Science Conference, {ITCS} 2017, January 9-11, 2017, Berkeley, CA,
  {USA}}}. \bibinfo{pages}{43:1--43:23}.
\newblock
\urldef\tempurl%
\url{https://doi.org/10.4230/LIPIcs.ITCS.2017.43}
\showDOI{\tempurl}


\bibitem[\protect\citeauthoryear{Knapp}{Knapp}{2016}]%
        {knapp_engaging_2016}
\bibfield{author}{\bibinfo{person}{Justin~Anthony Knapp}.}
  \bibinfo{year}{2016}\natexlab{}.
\newblock \showarticletitle{Engaging the {Public} in {Ethical} {Reasoning}
  {About} {Big} {Data}}.
\newblock In \bibinfo{booktitle}{\emph{Ethical {Reasoning} in {Big} {Data}:
  {An} {Exploratory} {Analysis}}}, \bibfield{editor}{\bibinfo{person}{Jeff
  Collman} {and} \bibinfo{person}{Sorin~Adam Matei}} (Eds.).
  \bibinfo{publisher}{Springer International Publishing}, \bibinfo{address}{New
  York, NY}, \bibinfo{pages}{43--52}.
\newblock
\showISBNx{978-3-319-28420-0}
\urldef\tempurl%
\url{https://doi.org/10.1007/978-3-319-28422-4_4}
\showURL{%
\tempurl}


\bibitem[\protect\citeauthoryear{Lage, Slavin~Ross, Kim, Gershman, and
  Doshi-Velez}{Lage et~al\mbox{.}}{2018}]%
        {lage_human---loop_2018}
\bibfield{author}{\bibinfo{person}{Isaac Lage}, \bibinfo{person}{Andrew
  Slavin~Ross}, \bibinfo{person}{Been Kim}, \bibinfo{person}{Samuel Gershman},
  {and} \bibinfo{person}{Finale Doshi-Velez}.} \bibinfo{year}{2018}\natexlab{}.
\newblock \showarticletitle{Human-in-the-{Loop} {Interpretability} {Prior}}. In
  \bibinfo{booktitle}{\emph{{NeurIPS} 2018}}. \bibinfo{address}{Montréal,
  Canada}.
\newblock
\urldef\tempurl%
\url{https://papers.nips.cc/paper/8219-human-in-the-loop-interpretability-prior.pdf}
\showURL{%
\tempurl}


\bibitem[\protect\citeauthoryear{Latour}{Latour}{1986}]%
        {latour_visualisation_1986}
\bibfield{author}{\bibinfo{person}{Bruno Latour}.}
  \bibinfo{year}{1986}\natexlab{}.
\newblock \showarticletitle{Visualisation and {Cognition}: {Thinking} with
  {Eyes} and {Hands}}.
\newblock In \bibinfo{booktitle}{\emph{Knowledge and {Society} {Studies} in the
  {Sociology} of {Culture} {Past} and {Present}}},
  \bibfield{editor}{\bibinfo{person}{H.~Kuklick}} (Ed.).
  Vol.~\bibinfo{volume}{6}. \bibinfo{publisher}{Jai Press},
  \bibinfo{pages}{1--40}.
\newblock


\bibitem[\protect\citeauthoryear{Lazar, Feng, and Hochheiser}{Lazar
  et~al\mbox{.}}{2010}]%
        {lazar_research_2010}
\bibfield{author}{\bibinfo{person}{Jonathan Lazar},
  \bibinfo{person}{Jinjuan~Heidi Feng}, {and} \bibinfo{person}{Harry
  Hochheiser}.} \bibinfo{year}{2010}\natexlab{}.
\newblock \bibinfo{booktitle}{\emph{Research methods in human-computer
  interaction}}.
\newblock \bibinfo{publisher}{John Wiley \& Sons},
  \bibinfo{address}{Indianapolis, IN}.
\newblock


\bibitem[\protect\citeauthoryear{Lehr and Ohm}{Lehr and Ohm}{2017}]%
        {LehrOhm2017}
\bibfield{author}{\bibinfo{person}{David Lehr} {and} \bibinfo{person}{Paul
  Ohm}.} \bibinfo{year}{2017}\natexlab{}.
\newblock \showarticletitle{Playing with the Data: What Legal Scholars Should
  Learn about Machine Learning}.
\newblock \bibinfo{journal}{\emph{UC Davis Law Review}} \bibinfo{volume}{51},
  \bibinfo{number}{2} (\bibinfo{year}{2017}), \bibinfo{pages}{653--717}.
\newblock


\bibitem[\protect\citeauthoryear{Leonelli}{Leonelli}{2016}]%
        {leonelli_locating_2016}
\bibfield{author}{\bibinfo{person}{Sabina Leonelli}.}
  \bibinfo{year}{2016}\natexlab{}.
\newblock \showarticletitle{Locating ethics in data science: responsibility and
  accountability in global and distributed knowledge production systems}.
\newblock \bibinfo{journal}{\emph{Philosophical transactions}}
  \bibinfo{volume}{374}, \bibinfo{number}{2083} (\bibinfo{year}{2016}),
  \bibinfo{pages}{20160122}.
\newblock
\urldef\tempurl%
\url{https://doi.org/10.1098/rsta.2016.0122}
\showDOI{\tempurl}


\bibitem[\protect\citeauthoryear{Lipton}{Lipton}{2016}]%
        {lipton_mythos_2016}
\bibfield{author}{\bibinfo{person}{Zachary Lipton}.}
  \bibinfo{year}{2016}\natexlab{}.
\newblock \showarticletitle{The mythos of model interpretability}.
\newblock \bibinfo{journal}{\emph{2016 ICML Workshop on Human Interpretability
  in Machine Learning (WHI 2016)}} (\bibinfo{year}{2016}).
\newblock


\bibitem[\protect\citeauthoryear{Lou, Caruana, and Gehrke}{Lou
  et~al\mbox{.}}{2012}]%
        {DBLP:conf/kdd/LouCG12}
\bibfield{author}{\bibinfo{person}{Yin Lou}, \bibinfo{person}{Rich Caruana},
  {and} \bibinfo{person}{Johannes Gehrke}.} \bibinfo{year}{2012}\natexlab{}.
\newblock \showarticletitle{Intelligible models for classification and
  regression}. In \bibinfo{booktitle}{\emph{The 18th {ACM} {SIGKDD}
  International Conference on Knowledge Discovery and Data Mining, {KDD} '12,
  Beijing, China, August 12-16, 2012}}. \bibinfo{pages}{150--158}.
\newblock
\urldef\tempurl%
\url{https://doi.org/10.1145/2339530.2339556}
\showDOI{\tempurl}


\bibitem[\protect\citeauthoryear{Luban, Strudler, and Wasserman}{Luban
  et~al\mbox{.}}{1992}]%
        {luban_moral_1992}
\bibfield{author}{\bibinfo{person}{David Luban}, \bibinfo{person}{Alan
  Strudler}, {and} \bibinfo{person}{David Wasserman}.}
  \bibinfo{year}{1992}\natexlab{}.
\newblock \showarticletitle{Moral {Responsibility} in the {Age} of
  {Bureaucracy}}.
\newblock \bibinfo{journal}{\emph{Michigan Law Review}} \bibinfo{volume}{90},
  \bibinfo{number}{8} (\bibinfo{year}{1992}), \bibinfo{pages}{2348--2392}.
\newblock


\bibitem[\protect\citeauthoryear{Lum and Isaac}{Lum and Isaac}{2016}]%
        {lum}
\bibfield{author}{\bibinfo{person}{Kristian Lum} {and} \bibinfo{person}{William
  Isaac}.} \bibinfo{year}{2016}\natexlab{}.
\newblock \showarticletitle{To predict and serve?}
\newblock \bibinfo{journal}{\emph{Significance}}  \bibinfo{volume}{13}
  (\bibinfo{year}{2016}), \bibinfo{pages}{14--19}.
\newblock
Issue 5.


\bibitem[\protect\citeauthoryear{Lundberg and Lee}{Lundberg and Lee}{2017}]%
        {lundberg_unified_2017}
\bibfield{author}{\bibinfo{person}{Scott Lundberg} {and} \bibinfo{person}{Su-In
  Lee}.} \bibinfo{year}{2017}\natexlab{}.
\newblock \showarticletitle{A {Unified} {Approach} to {Interpreting} {Model}
  {Predictions}}. In \bibinfo{booktitle}{\emph{Proceedings of {NIPS} 2017}}.
  \bibinfo{publisher}{NIPS}.
\newblock


\bibitem[\protect\citeauthoryear{Markham, Tiidenberg, and Herman}{Markham
  et~al\mbox{.}}{2018}]%
        {markham_ethics_2018}
\bibfield{author}{\bibinfo{person}{Annette~N. Markham}, \bibinfo{person}{Katrin
  Tiidenberg}, {and} \bibinfo{person}{Andrew Herman}.}
  \bibinfo{year}{2018}\natexlab{}.
\newblock \showarticletitle{Ethics as {Methods}: {Doing} {Ethics} in the {Era}
  of {Big} {Data} {Research}—{Introduction}}.
\newblock \bibinfo{journal}{\emph{Social Media + Society}} \bibinfo{volume}{4},
  \bibinfo{number}{3} (\bibinfo{year}{2018}), \bibinfo{pages}{1--9}.
\newblock
\urldef\tempurl%
\url{https://doi.org/10.1177/2056305118784502}
\showDOI{\tempurl}


\bibitem[\protect\citeauthoryear{Martin and Holz}{Martin and Holz}{1992}]%
        {martin_non-apologetic_1992}
\bibfield{author}{\bibinfo{person}{C.~Dianne Martin} {and}
  \bibinfo{person}{Hilary~J. Holz}.} \bibinfo{year}{1992}\natexlab{}.
\newblock \showarticletitle{Non-apologetic computer ethics education: a
  strategy for integrating social impact and ethics into the computer science
  curriculum}.
\newblock In \bibinfo{booktitle}{\emph{Teaching computer ethics}},
  \bibfield{editor}{\bibinfo{person}{Terrell~Ward Bynum},
  \bibinfo{person}{Walter Maner}, {and} \bibinfo{person}{John~L. Fodor}}
  (Eds.). \bibinfo{publisher}{Southern Connecticut State University},
  \bibinfo{address}{New Haven, CT}, \bibinfo{pages}{50--66}.
\newblock
\urldef\tempurl%
\url{https://rcvest.southernct.edu/teaching-computer-ethics/}
\showURL{%
\tempurl}


\bibitem[\protect\citeauthoryear{Mayer}{Mayer}{2010}]%
        {mayer_applying_2010}
\bibfield{author}{\bibinfo{person}{Richard~E. Mayer}.}
  \bibinfo{year}{2010}\natexlab{}.
\newblock \bibinfo{booktitle}{\emph{Applying the {Science} of {Learning}}}.
\newblock \bibinfo{publisher}{Pearson}, \bibinfo{address}{New York, NY}.
\newblock
\showISBNx{978-0136117575}


\bibitem[\protect\citeauthoryear{Mervis}{Mervis}{2019}]%
        {mervis_can_2019}
\bibfield{author}{\bibinfo{person}{Jeffrey Mervis}.}
  \bibinfo{year}{2019}\natexlab{}.
\newblock \showarticletitle{Can a set of equations keep {U}.{S}. census data
  private?}
\newblock \bibinfo{journal}{\emph{Science Magazine}} (\bibinfo{date}{Jan.}
  \bibinfo{year}{2019}).
\newblock


\bibitem[\protect\citeauthoryear{Mitchell, Potash, Barocas, D'Amour, and
  Lum}{Mitchell et~al\mbox{.}}{2018}]%
        {mitchell}
\bibfield{author}{\bibinfo{person}{Shira Mitchell}, \bibinfo{person}{Eric
  Potash}, \bibinfo{person}{Solon Barocas}, \bibinfo{person}{Alexander
  D'Amour}, {and} \bibinfo{person}{Kristian Lum}.}
  \bibinfo{year}{2018}\natexlab{}.
\newblock \showarticletitle{Prediction-Based Decisions and Fairness: A
  Catalogue of Choices, Assumptions, and Definitions}.
\newblock \bibinfo{journal}{\emph{CoRR}}  \bibinfo{volume}{abs/1811.07867}
  (\bibinfo{year}{2018}).
\newblock
\urldef\tempurl%
\url{https://arxiv.org/abs/1811.07867}
\showURL{%
\tempurl}


\bibitem[\protect\citeauthoryear{Moreau, Lud{\"{a}}scher, Altintas, Barga,
  Bowers, Callahan, Jr., Clifford, Cohen, Boulakia, Davidson, Deelman,
  Digiampietri, Foster, Freire, Frew, Futrelle, Gibson, Gil, Goble, Golbeck,
  Groth, Holland, Jiang, Kim, Koop, Krenek, McPhillips, Mehta, Miles, Metzger,
  Munroe, Myers, Plale, Podhorszki, Ratnakar, Santos, Scheidegger, Schuchardt,
  Seltzer, Simmhan, Silva, Slaughter, Stephan, Stevens, Turi, Vo, Wilde, Zhao,
  and Zhao}{Moreau et~al\mbox{.}}{2008}]%
        {DBLP:journals/concurrency/MoreauLA08}
\bibfield{author}{\bibinfo{person}{Luc Moreau}, \bibinfo{person}{Bertram
  Lud{\"{a}}scher}, \bibinfo{person}{Ilkay Altintas}, \bibinfo{person}{Roger~S.
  Barga}, \bibinfo{person}{Shawn Bowers}, \bibinfo{person}{Steven~P. Callahan},
  \bibinfo{person}{George~Chin Jr.}, \bibinfo{person}{Ben Clifford},
  \bibinfo{person}{Shirley Cohen}, \bibinfo{person}{Sarah~Cohen Boulakia},
  \bibinfo{person}{Susan~B. Davidson}, \bibinfo{person}{Ewa Deelman},
  \bibinfo{person}{Luciano~A. Digiampietri}, \bibinfo{person}{Ian~T. Foster},
  \bibinfo{person}{Juliana Freire}, \bibinfo{person}{James Frew},
  \bibinfo{person}{Joe Futrelle}, \bibinfo{person}{Tara Gibson},
  \bibinfo{person}{Yolanda Gil}, \bibinfo{person}{Carole~A. Goble},
  \bibinfo{person}{Jennifer Golbeck}, \bibinfo{person}{Paul~T. Groth},
  \bibinfo{person}{David~A. Holland}, \bibinfo{person}{Sheng Jiang},
  \bibinfo{person}{Jihie Kim}, \bibinfo{person}{David Koop},
  \bibinfo{person}{Ales Krenek}, \bibinfo{person}{Timothy~M. McPhillips},
  \bibinfo{person}{Gaurang Mehta}, \bibinfo{person}{Simon Miles},
  \bibinfo{person}{Dominic Metzger}, \bibinfo{person}{Steve Munroe},
  \bibinfo{person}{Jim Myers}, \bibinfo{person}{Beth Plale},
  \bibinfo{person}{Norbert Podhorszki}, \bibinfo{person}{Varun Ratnakar},
  \bibinfo{person}{Emanuele Santos}, \bibinfo{person}{Carlos~Eduardo
  Scheidegger}, \bibinfo{person}{Karen Schuchardt}, \bibinfo{person}{Margo~I.
  Seltzer}, \bibinfo{person}{Yogesh~L. Simmhan},
  \bibinfo{person}{Cl{\'{a}}udio~T. Silva}, \bibinfo{person}{Peter Slaughter},
  \bibinfo{person}{Eric~G. Stephan}, \bibinfo{person}{Robert Stevens},
  \bibinfo{person}{Daniele Turi}, \bibinfo{person}{Huy~T. Vo},
  \bibinfo{person}{Michael Wilde}, \bibinfo{person}{Jun Zhao}, {and}
  \bibinfo{person}{Yong Zhao}.} \bibinfo{year}{2008}\natexlab{}.
\newblock \showarticletitle{Special Issue: The First Provenance Challenge}.
\newblock \bibinfo{journal}{\emph{Concurrency and Computation: Practice and
  Experience}} \bibinfo{volume}{20}, \bibinfo{number}{5}
  (\bibinfo{year}{2008}), \bibinfo{pages}{409--418}.
\newblock
\urldef\tempurl%
\url{https://doi.org/10.1002/cpe.1233}
\showDOI{\tempurl}


\bibitem[\protect\citeauthoryear{Mumford, Connelly, Brown, Murphy, Hill, Antes,
  Waples, and Devenport}{Mumford et~al\mbox{.}}{2008}]%
        {mumford_sensemaking_2008}
\bibfield{author}{\bibinfo{person}{Michael~D. Mumford}, \bibinfo{person}{Shane
  Connelly}, \bibinfo{person}{Ryan~P. Brown}, \bibinfo{person}{Stephen~T.
  Murphy}, \bibinfo{person}{Jason~H. Hill}, \bibinfo{person}{Alison~L. Antes},
  \bibinfo{person}{Ethan~P. Waples}, {and} \bibinfo{person}{Lynn~D.
  Devenport}.} \bibinfo{year}{2008}\natexlab{}.
\newblock \showarticletitle{A {Sensemaking} {Approach} to {Ethics} {Training}
  for {Scientists}: {Preliminary} {Evidence} of {Training} {Effectiveness}}.
\newblock \bibinfo{journal}{\emph{Ethics \& Behavior}} \bibinfo{volume}{18},
  \bibinfo{number}{4} (\bibinfo{year}{2008}), \bibinfo{pages}{315--339}.
\newblock
\showISSN{1050-8422}
\urldef\tempurl%
\url{https://doi.org/10.1080/10508420802487815}
\showDOI{\tempurl}


\bibitem[\protect\citeauthoryear{Naps, Rößling, Almstrum, Dann, Fleischer,
  Hundhausen, Korhonen, Malmi, McNally, Rodger, and Velázquez-Iturbide}{Naps
  et~al\mbox{.}}{2002}]%
        {naps_exploring_2002}
\bibfield{author}{\bibinfo{person}{Thomas~L. Naps}, \bibinfo{person}{Guido
  Rößling}, \bibinfo{person}{Vicki Almstrum}, \bibinfo{person}{Wanda Dann},
  \bibinfo{person}{Rudolf Fleischer}, \bibinfo{person}{Christopher Hundhausen},
  \bibinfo{person}{Ari Korhonen}, \bibinfo{person}{Lauri Malmi},
  \bibinfo{person}{Myles McNally}, \bibinfo{person}{Susan Rodger}, {and}
  \bibinfo{person}{J.~Ángel Velázquez-Iturbide}.}
  \bibinfo{year}{2002}\natexlab{}.
\newblock \showarticletitle{Exploring the role of visualization and engagement
  in computer science education}. In \bibinfo{booktitle}{\emph{Proceedings of
  {ITiCSE}-{WGR} '02}}. \bibinfo{publisher}{ACM}, \bibinfo{address}{Aarhus,
  Denmark}, \bibinfo{pages}{131--152}.
\newblock


\bibitem[\protect\citeauthoryear{Nolan and Perrett}{Nolan and Perrett}{2016}]%
        {nolan_teaching_2016}
\bibfield{author}{\bibinfo{person}{Deborah Nolan} {and} \bibinfo{person}{Jamis
  Perrett}.} \bibinfo{year}{2016}\natexlab{}.
\newblock \showarticletitle{Teaching and {Learning} {Data} {Visualization}:
  {Ideas} and {Assignments}}.
\newblock \bibinfo{journal}{\emph{The American Statistician}}
  \bibinfo{volume}{70}, \bibinfo{number}{3} (\bibinfo{year}{2016}),
  \bibinfo{pages}{260--269}.
\newblock


\bibitem[\protect\citeauthoryear{O'Neil}{O'Neil}{2016}]%
        {oneil_weapons_2016}
\bibfield{author}{\bibinfo{person}{Cathy O'Neil}.}
  \bibinfo{year}{2016}\natexlab{}.
\newblock \bibinfo{booktitle}{\emph{Weapons of {Math} {Destruction}: {How}
  {Big} {Data} {Increases} {Inequality} and {Threatens} {Democracy}}}.
\newblock \bibinfo{publisher}{Crown}, \bibinfo{address}{New York, NY}.
\newblock
\showISBNx{978-0553418811}


\bibitem[\protect\citeauthoryear{Papert}{Papert}{1980}]%
        {papert_mindstorms:_1980}
\bibfield{author}{\bibinfo{person}{Seymour Papert}.}
  \bibinfo{year}{1980}\natexlab{}.
\newblock \bibinfo{booktitle}{\emph{Mindstorms: {Children}, computers, and
  powerful ideas}}.
\newblock \bibinfo{publisher}{Basic Books}, \bibinfo{address}{New York, NY}.
\newblock


\bibitem[\protect\citeauthoryear{Passi and Barocas}{Passi and Barocas}{2019}]%
        {passi_problem_2019}
\bibfield{author}{\bibinfo{person}{Samir Passi} {and} \bibinfo{person}{Solon
  Barocas}.} \bibinfo{year}{2019}\natexlab{}.
\newblock \showarticletitle{Problem {Formulation} and {Fairness}}. In
  \bibinfo{booktitle}{\emph{Proceedings of the {ACM} {Conference} on
  {Fairness}, {Accountability}, and {Transparency}}}. \bibinfo{publisher}{ACM},
  \bibinfo{pages}{10}.
\newblock
\urldef\tempurl%
\url{https://arxiv.org/pdf/1901.02547.pdf}
\showURL{%
\tempurl}


\bibitem[\protect\citeauthoryear{Ping, Stoyanovich, and Howe}{Ping
  et~al\mbox{.}}{2017}]%
        {DBLP:conf/ssdbm/PingSH17}
\bibfield{author}{\bibinfo{person}{Haoyue Ping}, \bibinfo{person}{Julia
  Stoyanovich}, {and} \bibinfo{person}{Bill Howe}.}
  \bibinfo{year}{2017}\natexlab{}.
\newblock \showarticletitle{DataSynthesizer: Privacy-Preserving Synthetic
  Datasets}. In \bibinfo{booktitle}{\emph{Proceedings of the 29th International
  Conference on Scientific and Statistical Database Management, Chicago, IL,
  USA, June 27-29, 2017}}. \bibinfo{pages}{42:1--42:5}.
\newblock
\urldef\tempurl%
\url{https://doi.org/10.1145/3085504.3091117}
\showDOI{\tempurl}


\bibitem[\protect\citeauthoryear{Poursabzi-Sangdeh, Vaughan, Goldstein, Hofman,
  and Wallach}{Poursabzi-Sangdeh et~al\mbox{.}}{2018}]%
        {poursabzi-sangdeh_manipulating_2018}
\bibfield{author}{\bibinfo{person}{Forough Poursabzi-Sangdeh},
  \bibinfo{person}{Jennifer~Wortman Vaughan}, \bibinfo{person}{Daniel~G.
  Goldstein}, \bibinfo{person}{Jake~M. Hofman}, {and} \bibinfo{person}{Hannah
  Wallach}.} \bibinfo{year}{2018}\natexlab{}.
\newblock \showarticletitle{Manipulating and {Measuring} {Model}
  {Interpretability}}.
\newblock  (\bibinfo{year}{2018}).
\newblock
\urldef\tempurl%
\url{https://arxiv.org/pdf/1802.07810.pdf}
\showURL{%
\tempurl}


\bibitem[\protect\citeauthoryear{Provost and Fawcett}{Provost and
  Fawcett}{2013}]%
        {provost_data_2013}
\bibfield{author}{\bibinfo{person}{Foster Provost} {and} \bibinfo{person}{Tom
  Fawcett}.} \bibinfo{year}{2013}\natexlab{}.
\newblock \bibinfo{booktitle}{\emph{Data {Science} for {Business}: {What} {You}
  {Need} to {Know} {About} {Data} {Mining} and {Data}-{Analytic} {Thinking}}}.
\newblock \bibinfo{publisher}{O’Reilly Media}, \bibinfo{address}{Sebastopol,
  CA}.
\newblock


\bibitem[\protect\citeauthoryear{Quinn}{Quinn}{2006}]%
        {quinn_teaching_2006}
\bibfield{author}{\bibinfo{person}{Michael~J. Quinn}.}
  \bibinfo{year}{2006}\natexlab{}.
\newblock \showarticletitle{On teaching computer ethics within a computer
  science department}.
\newblock \bibinfo{journal}{\emph{Science and Engineering Ethics}}
  \bibinfo{volume}{12}, \bibinfo{number}{2} (\bibinfo{date}{June}
  \bibinfo{year}{2006}), \bibinfo{pages}{335--343}.
\newblock
\showISSN{1471-5546}
\urldef\tempurl%
\url{https://doi.org/10.1007/s11948-006-0032-9}
\showDOI{\tempurl}


\bibitem[\protect\citeauthoryear{Ribeiro, Singh, and Guestrin}{Ribeiro
  et~al\mbox{.}}{2016}]%
        {ribeiro_why_2016}
\bibfield{author}{\bibinfo{person}{Marco~Tulio Ribeiro},
  \bibinfo{person}{Sameer Singh}, {and} \bibinfo{person}{Carlos Guestrin}.}
  \bibinfo{year}{2016}\natexlab{}.
\newblock \showarticletitle{”{Why} {Should} {I} {Trust} {You}?”:
  {Explaining} the predictions of any classifier}. In
  \bibinfo{booktitle}{\emph{{KDD} '16 {Proceedings} of the 22nd {ACM} {SIGKDD}
  {International} {Conference} on {Knowledge} {Discovery} and {Data}
  {Mining}}}. \bibinfo{publisher}{ACM}, \bibinfo{address}{San Francisco, CA,
  USA}, \bibinfo{pages}{1135--1144}.
\newblock
\urldef\tempurl%
\url{https://doi.org/10.1145/2939672.2939778}
\showDOI{\tempurl}


\bibitem[\protect\citeauthoryear{Rubin, Hammerman, and Konold}{Rubin
  et~al\mbox{.}}{2006}]%
        {rubin_exploring_2006}
\bibfield{author}{\bibinfo{person}{Andee Rubin}, \bibinfo{person}{James
  Hammerman}, {and} \bibinfo{person}{Cliff Konold}.}
  \bibinfo{year}{2006}\natexlab{}.
\newblock \showarticletitle{Exploring informal inference with interactive
  visualization software}. In \bibinfo{booktitle}{\emph{Proceedings of the
  {Seventh} {International} {Conference} on {Teaching} {Statistics}}}.
\newblock


\bibitem[\protect\citeauthoryear{Salganik}{Salganik}{2019}]%
        {salganik}
\bibfield{author}{\bibinfo{person}{Matthew~J. Salganik}.}
  \bibinfo{year}{2019}\natexlab{}.
\newblock \bibinfo{booktitle}{\emph{Bit By Bit: Social Research in the Digital
  Age}}.
\newblock \bibinfo{publisher}{Princeton University Press}.
\newblock


\bibitem[\protect\citeauthoryear{Salimi, Rodriguez, Howe, and Suciu}{Salimi
  et~al\mbox{.}}{2019}]%
        {DBLP:conf/sigmod/SalimiRHS19}
\bibfield{author}{\bibinfo{person}{Babak Salimi}, \bibinfo{person}{Luke
  Rodriguez}, \bibinfo{person}{Bill Howe}, {and} \bibinfo{person}{Dan Suciu}.}
  \bibinfo{year}{2019}\natexlab{}.
\newblock \showarticletitle{Interventional Fairness: Causal Database Repair for
  Algorithmic Fairness}. In \bibinfo{booktitle}{\emph{Proceedings of the 2019
  International Conference on Management of Data, {SIGMOD} Conference 2019,
  Amsterdam, The Netherlands, June 30 - July 5, 2019.}}
  \bibinfo{pages}{793--810}.
\newblock
\urldef\tempurl%
\url{https://doi.org/10.1145/3299869.3319901}
\showDOI{\tempurl}


\bibitem[\protect\citeauthoryear{Schraagen, Chipman, and Shalin}{Schraagen
  et~al\mbox{.}}{2000}]%
        {schraagen_introduction_2000}
\bibfield{author}{\bibinfo{person}{Jan~Maarten Schraagen},
  \bibinfo{person}{Susan~F. Chipman}, {and} \bibinfo{person}{Valerie~L.
  Shalin}.} \bibinfo{year}{2000}\natexlab{}.
\newblock \showarticletitle{Introduction to cognitive task analysis}.
\newblock In \bibinfo{booktitle}{\emph{Cognitive task analysis}},
  \bibfield{editor}{\bibinfo{person}{Jan~Maarten Schraagen},
  \bibinfo{person}{Susan~F. Chipman}, {and} \bibinfo{person}{Valerie~L.
  Shalin}} (Eds.). \bibinfo{publisher}{Erlbaum}, \bibinfo{address}{Mahwah, NJ},
  \bibinfo{pages}{3--23}.
\newblock


\bibitem[\protect\citeauthoryear{Selbst and Barocas}{Selbst and
  Barocas}{2018}]%
        {selbst_intuitive_2018}
\bibfield{author}{\bibinfo{person}{Andrew Selbst} {and} \bibinfo{person}{Solon
  Barocas}.} \bibinfo{year}{2018}\natexlab{}.
\newblock \showarticletitle{The {Intuitive} {Appeal} of {Explainable}
  {Machines}}.
\newblock \bibinfo{journal}{\emph{Fordham Law Review}} \bibinfo{volume}{87},
  \bibinfo{number}{3} (\bibinfo{year}{2018}), \bibinfo{pages}{1085 --1139}.
\newblock


\bibitem[\protect\citeauthoryear{Shmueli}{Shmueli}{2010}]%
        {shmueli_explain_2010}
\bibfield{author}{\bibinfo{person}{Galit Shmueli}.}
  \bibinfo{year}{2010}\natexlab{}.
\newblock \showarticletitle{To explain or to predict?}
\newblock \bibinfo{journal}{\emph{Statist. Sci.}} \bibinfo{volume}{25},
  \bibinfo{number}{3} (\bibinfo{year}{2010}), \bibinfo{pages}{289--310}.
\newblock
\urldef\tempurl%
\url{https://doi.org/10.1214/10-STS330}
\showDOI{\tempurl}


\bibitem[\protect\citeauthoryear{Sriadhi, Rahim, and Ahmar}{Sriadhi
  et~al\mbox{.}}{2018}]%
        {sriadhi_rc4_2018}
\bibfield{author}{\bibinfo{person}{S. Sriadhi}, \bibinfo{person}{Robbi Rahim},
  {and} \bibinfo{person}{Ansari~Saleh Ahmar}.} \bibinfo{year}{2018}\natexlab{}.
\newblock \showarticletitle{{RC}4 {Algorithm} {Visualization} for
  {Cryptography} {Education}}.
\newblock \bibinfo{journal}{\emph{Journal of Physics: Conference Series}}
  \bibinfo{volume}{1028} (\bibinfo{year}{2018}).
\newblock
\urldef\tempurl%
\url{https://doi.org/10.1088/1742-6596/1028/1/012057}
\showDOI{\tempurl}


\bibitem[\protect\citeauthoryear{Sternberg}{Sternberg}{2010}]%
        {sternberg_teaching_2010}
\bibfield{author}{\bibinfo{person}{Robert~J. Sternberg}.}
  \bibinfo{year}{2010}\natexlab{}.
\newblock \showarticletitle{Teaching for {Ethical} {Reasoning} in {Liberal}
  {Education}}.
\newblock \bibinfo{journal}{\emph{Liberal Education (Association of American
  Colleges \& Universities)}} \bibinfo{volume}{96}, \bibinfo{number}{3}
  (\bibinfo{year}{2010}).
\newblock
\urldef\tempurl%
\url{https://www.aacu.org/publications-research/periodicals/teaching-ethical-reasoning-liberal-education}
\showURL{%
\tempurl}


\bibitem[\protect\citeauthoryear{Stoyanovich and Goodman}{Stoyanovich and
  Goodman}{2016}]%
        {label}
\bibfield{author}{\bibinfo{person}{Julia Stoyanovich} {and}
  \bibinfo{person}{Ellen~P. Goodman}.} \bibinfo{year}{2016}\natexlab{}.
\newblock \showarticletitle{Revealing Algorithmic Rankers}.
\newblock
  \bibinfo{howpublished}{\url{http://freedom-to-tinker.com/2016/08/05/revealing-algorithmic-rankers/}}.
\newblock \bibinfo{journal}{\emph{Freedom to Tinker}} (\bibinfo{date}{August 5}
  \bibinfo{year}{2016}).
\newblock
\newblock
\shownote{[Online; accessed 19-August-2019].}


\bibitem[\protect\citeauthoryear{Stoyanovich and Howe}{Stoyanovich and
  Howe}{2018a}]%
        {follow}
\bibfield{author}{\bibinfo{person}{Julia Stoyanovich} {and}
  \bibinfo{person}{Bill Howe}.} \bibinfo{year}{2018}\natexlab{a}.
\newblock \showarticletitle{Follow the data! Algorithmic transparency starts
  with data transparency}.
\newblock
  \bibinfo{howpublished}{\url{https://ai.shorensteincenter.org/ideas/2018/11/26/follow-the-data-algorithmic-transparency-starts-with-data-transparency}}.
\newblock \bibinfo{journal}{\emph{The Ethical Machine}}
  (\bibinfo{date}{November 27} \bibinfo{year}{2018}).
\newblock
\newblock
\shownote{[Online; accessed 19-August-2019].}


\bibitem[\protect\citeauthoryear{Stoyanovich and Howe}{Stoyanovich and
  Howe}{2018b}]%
        {refining}
\bibfield{author}{\bibinfo{person}{Julia Stoyanovich} {and}
  \bibinfo{person}{Bill Howe}.} \bibinfo{year}{2018}\natexlab{b}.
\newblock \showarticletitle{Refining the Concept of a Nutritional Label for
  Data and Models}.
\newblock
  \bibinfo{howpublished}{\url{https://freedom-to-tinker.com/2018/05/03/refining-the-concept-of-a-nutritional-label-for-data-and-models/}}.
\newblock \bibinfo{journal}{\emph{Freedom to Tinker}} (\bibinfo{date}{May 3}
  \bibinfo{year}{2018}).
\newblock
\newblock
\shownote{[Online; accessed 19-August-2019].}


\bibitem[\protect\citeauthoryear{Stoyanovich, Howe, Abiteboul, Miklau,
  Sahuguet, and Weikum}{Stoyanovich et~al\mbox{.}}{2017}]%
        {DBLP:conf/ssdbm/StoyanovichHAMS17}
\bibfield{author}{\bibinfo{person}{Julia Stoyanovich}, \bibinfo{person}{Bill
  Howe}, \bibinfo{person}{Serge Abiteboul}, \bibinfo{person}{Gerome Miklau},
  \bibinfo{person}{Arnaud Sahuguet}, {and} \bibinfo{person}{Gerhard Weikum}.}
  \bibinfo{year}{2017}\natexlab{}.
\newblock \showarticletitle{Fides: Towards a Platform for Responsible Data
  Science}. In \bibinfo{booktitle}{\emph{Proceedings of the 29th International
  Conference on Scientific and Statistical Database Management, Chicago, IL,
  USA, June 27-29, 2017}}. \bibinfo{pages}{26:1--26:6}.
\newblock
\urldef\tempurl%
\url{https://doi.org/10.1145/3085504.3085530}
\showDOI{\tempurl}


\bibitem[\protect\citeauthoryear{Stoyanovich, Yang, and Jagadish}{Stoyanovich
  et~al\mbox{.}}{2018}]%
        {DBLP:conf/edbt/StoyanovichYJ18}
\bibfield{author}{\bibinfo{person}{Julia Stoyanovich}, \bibinfo{person}{Ke
  Yang}, {and} \bibinfo{person}{H.~V. Jagadish}.}
  \bibinfo{year}{2018}\natexlab{}.
\newblock \showarticletitle{Online Set Selection with Fairness and Diversity
  Constraints}. In \bibinfo{booktitle}{\emph{Proceedings of the 21th
  International Conference on Extending Database Technology, {EDBT} 2018,
  Vienna, Austria, March 26-29, 2018.}} \bibinfo{pages}{241--252}.
\newblock
\urldef\tempurl%
\url{https://doi.org/10.5441/002/edbt.2018.22}
\showDOI{\tempurl}


\bibitem[\protect\citeauthoryear{Sweeney}{Sweeney}{2013}]%
        {DBLP:journals/queue/Sweeney13}
\bibfield{author}{\bibinfo{person}{Latanya Sweeney}.}
  \bibinfo{year}{2013}\natexlab{}.
\newblock \showarticletitle{Discrimination in Online Ad Delivery}.
\newblock \bibinfo{journal}{\emph{{ACM} Queue}} \bibinfo{volume}{11},
  \bibinfo{number}{3} (\bibinfo{year}{2013}), \bibinfo{pages}{10}.
\newblock
\urldef\tempurl%
\url{https://doi.org/10.1145/2460276.2460278}
\showDOI{\tempurl}


\bibitem[\protect\citeauthoryear{Tomlinson}{Tomlinson}{2017}]%
        {tomlinson_how_2017}
\bibfield{author}{\bibinfo{person}{Carol~Ann Tomlinson}.}
  \bibinfo{year}{2017}\natexlab{}.
\newblock \bibinfo{booktitle}{\emph{How to {Differentiate} {Instruction} in
  {Academically} {Diverse} {Classroom}} (\bibinfo{edition}{3rd} ed.)}.
\newblock \bibinfo{publisher}{ASCD}, \bibinfo{address}{Alexandria, VA}.
\newblock
\showISBNx{978-1-4166-2330-4}


\bibitem[\protect\citeauthoryear{Tractenberg, Russell, Morgan, FitzGerald,
  Collmann, Vinsel, Steinmann, and Dolling}{Tractenberg et~al\mbox{.}}{2015}]%
        {tractenberg_using_2015}
\bibfield{author}{\bibinfo{person}{Rochelle~E. Tractenberg},
  \bibinfo{person}{Andrew~J. Russell}, \bibinfo{person}{Gregory~J. Morgan},
  \bibinfo{person}{Kevin~T. FitzGerald}, \bibinfo{person}{Jeff Collmann},
  \bibinfo{person}{Lee Vinsel}, \bibinfo{person}{Michael Steinmann}, {and}
  \bibinfo{person}{Lisa~M. Dolling}.} \bibinfo{year}{2015}\natexlab{}.
\newblock \showarticletitle{Using {Ethical} {Reasoning} to {Amplify} the
  {Reach} and {Resonance} of {Professional} {Codes} of {Conduct} in {Training}
  {Big} {Data} {Scientists}}.
\newblock \bibinfo{journal}{\emph{Science and Engineering Ethics}}
  \bibinfo{volume}{21}, \bibinfo{number}{6} (\bibinfo{year}{2015}),
  \bibinfo{pages}{1485--1507}.
\newblock
\urldef\tempurl%
\url{https://doi.org/10.1007/s11948-014-9613-1}
\showDOI{\tempurl}


\bibitem[\protect\citeauthoryear{Tufte}{Tufte}{2001}]%
        {tufte_visual_2001}
\bibfield{author}{\bibinfo{person}{Edward~R. Tufte}.}
  \bibinfo{year}{2001}\natexlab{}.
\newblock \bibinfo{booktitle}{\emph{The {Visual} {Display} of {Quantitative}
  {Information}} (\bibinfo{edition}{second} ed.)}.
\newblock \bibinfo{publisher}{Graphics Press}, \bibinfo{address}{Cheshire, CT}.
\newblock


\bibitem[\protect\citeauthoryear{Tukey}{Tukey}{1977}]%
        {tukey_exploratory_1977}
\bibfield{author}{\bibinfo{person}{John~W. Tukey}.}
  \bibinfo{year}{1977}\natexlab{}.
\newblock \bibinfo{booktitle}{\emph{Exploratory data analysis}}.
\newblock \bibinfo{publisher}{Addison-Wesley}, \bibinfo{address}{Reading, PA}.
\newblock


\bibitem[\protect\citeauthoryear{Turkle and Papert}{Turkle and Papert}{1992}]%
        {turkle_epistemological_1992}
\bibfield{author}{\bibinfo{person}{Sherry Turkle} {and}
  \bibinfo{person}{Seymour Papert}.} \bibinfo{year}{1992}\natexlab{}.
\newblock \showarticletitle{Epistemological {Pluralism} and the {Revaluation}
  of the {Concrete}}.
\newblock \bibinfo{journal}{\emph{Journal of Mathematical Behavior}}
  \bibinfo{volume}{11}, \bibinfo{number}{1} (\bibinfo{date}{March}
  \bibinfo{year}{1992}), \bibinfo{pages}{3--33}.
\newblock


\bibitem[\protect\citeauthoryear{University}{University}{2019}]%
        {harvard_university_embedded_2019}
\bibfield{author}{\bibinfo{person}{Harvard University}.}
  \bibinfo{year}{2019}\natexlab{}.
\newblock \bibinfo{title}{Embedded {EthiCS} @ {Harvard}}.
\newblock
\newblock
\urldef\tempurl%
\url{https://embeddedethics.seas.harvard.edu/about.html}
\showURL{%
\tempurl}


\bibitem[\protect\citeauthoryear{Valentino-DeVries, Singer-Vine, and
  Soltani}{Valentino-DeVries et~al\mbox{.}}{2012}]%
        {staples}
\bibfield{author}{\bibinfo{person}{Jennifer Valentino-DeVries},
  \bibinfo{person}{Jeremy Singer-Vine}, {and} \bibinfo{person}{Ashkan
  Soltani}.} \bibinfo{year}{2012}\natexlab{}.
\newblock \bibinfo{title}{Websites Vary Prices, Deals Based on Users'
  Information}.
\newblock
\newblock
\urldef\tempurl%
\url{https://www.wsj.com/articles/SB10001424127887323777204578189391813881534}
\showURL{%
\tempurl}


\bibitem[\protect\citeauthoryear{Wenger}{Wenger}{1998}]%
        {wenger_communities_1998}
\bibfield{author}{\bibinfo{person}{Etienne Wenger}.}
  \bibinfo{year}{1998}\natexlab{}.
\newblock \bibinfo{booktitle}{\emph{Communities of practice}}.
\newblock \bibinfo{publisher}{Cambridge University Press},
  \bibinfo{address}{Cambridge, England}.
\newblock


\bibitem[\protect\citeauthoryear{Wierse and Grinstein}{Wierse and
  Grinstein}{2002}]%
        {fayyad_information_2002}
\bibfield{author}{\bibinfo{person}{Andreas Wierse} {and}
  \bibinfo{person}{Georges Grinstein}.} \bibinfo{year}{2002}\natexlab{}.
\newblock \bibinfo{booktitle}{\emph{Information {Visualization} in {Data}
  {Mining} and {Knowledge} {Discovery}}}.
\newblock \bibinfo{publisher}{Morgan Kauffmann Publishers},
  \bibinfo{address}{San Francisco, CA}.
\newblock


\bibitem[\protect\citeauthoryear{Williamson}{Williamson}{2016}]%
        {williamson_digital_2016}
\bibfield{author}{\bibinfo{person}{Ben Williamson}.}
  \bibinfo{year}{2016}\natexlab{}.
\newblock \showarticletitle{Digital education governance: data visualization,
  predictive analytics, and ‘real-time’ policy instruments}.
\newblock \bibinfo{journal}{\emph{Journal of Education Policy}}
  \bibinfo{volume}{31}, \bibinfo{number}{2} (\bibinfo{year}{2016}),
  \bibinfo{pages}{123--141}.
\newblock
\urldef\tempurl%
\url{https://doi.org/10.1080/02680939.2015.1035758}
\showDOI{\tempurl}


\bibitem[\protect\citeauthoryear{Yang and Stoyanovich}{Yang and
  Stoyanovich}{2017}]%
        {DBLP:conf/ssdbm/YangS17}
\bibfield{author}{\bibinfo{person}{Ke Yang} {and} \bibinfo{person}{Julia
  Stoyanovich}.} \bibinfo{year}{2017}\natexlab{}.
\newblock \showarticletitle{Measuring Fairness in Ranked Outputs}. In
  \bibinfo{booktitle}{\emph{Proceedings of the 29th International Conference on
  Scientific and Statistical Database Management, Chicago, IL, USA, June 27-29,
  2017}}. \bibinfo{pages}{22:1--22:6}.
\newblock
\urldef\tempurl%
\url{https://doi.org/10.1145/3085504.3085526}
\showDOI{\tempurl}


\bibitem[\protect\citeauthoryear{Yang, Stoyanovich, Asudeh, Howe, Jagadish, and
  Miklau}{Yang et~al\mbox{.}}{2018}]%
        {DBLP:conf/sigmod/YangSAHJM18}
\bibfield{author}{\bibinfo{person}{Ke Yang}, \bibinfo{person}{Julia
  Stoyanovich}, \bibinfo{person}{Abolfazl Asudeh}, \bibinfo{person}{Bill Howe},
  \bibinfo{person}{H.~V. Jagadish}, {and} \bibinfo{person}{Gerome Miklau}.}
  \bibinfo{year}{2018}\natexlab{}.
\newblock \showarticletitle{A Nutritional Label for Rankings}. In
  \bibinfo{booktitle}{\emph{Proceedings of the 2018 International Conference on
  Management of Data, {SIGMOD} Conference 2018, Houston, TX, USA, June 10-15,
  2018}}. \bibinfo{pages}{1773--1776}.
\newblock
\urldef\tempurl%
\url{https://doi.org/10.1145/3183713.3193568}
\showDOI{\tempurl}


\end{thebibliography}
